\documentclass[10pt,reqno,letterpaper]{article}
\usepackage{makeidx}
\makeindex
\usepackage{geometry}              
\usepackage{graphicx}
\usepackage[dvipdfmx]{color}
\usepackage{amssymb}
\usepackage{amsmath}
\usepackage{setspace}
\usepackage{ascmac}
\usepackage{wrapfig}
\usepackage{fancybox}

\newcommand{\Integer}{{\mathbb{Z}}}

\newcommand{\Split}{\begin{split}}
\newcommand{\SplitX}{\end{split}}
\newcommand{\Gather}{\begin{gather}}
\newcommand{\GatherX}{\end{gather}}

\newcommand{\eqfig}[1]{\raisebox{-10pt}{\includegraphics[scale=0.8]{#1}}}

\begin{document}

%
%
\begin{titlepage}
\begin{flushright}
\normalsize
~~~~
OCU-PHYS 407\\
July, 2014\\
\end{flushright}

\vspace{15pt}

\begin{center}
{\LARGE  Exchange Relation in $sl_{3}$ WZNW model}\\~\\
{\LARGE in Semiclassical Limit}\\~\\
\end{center}

\vspace{23pt}

\begin{center}
{ Sho Deguchi$^{a}$\footnote{e-mail: deguchi@sci.osaka-cu.ac.jp}
}\\
%
\vspace{18pt}
%

$^a$ \it Department of Mathematics and Physics, Graduate School of Science\\
Osaka City University\\

\vspace{5pt}

3-3-138, Sugimoto, Sumiyoshi-ku, Osaka, 558-8585, Japan \\

\end{center}
%
\vspace{20pt}
\begin{center}
Abstract\\
\end{center}
We consider the exchange relations of screened vertex operators in the $sl_{3}$ Wess-Zumino-Novikov-Witten(WZNW) model in the semiclassical limit (where level $k$ tends to infinity).
We demonstrate that the coefficients of the exchange relations of the screened vertex operators coincide with the ones of the spider diagrams\cite{Archer:1992kc,1997q.alg....12003K}. 
The spider diagrams are composed of $3$-point vertices, and differ from ordinary string diagrams.


\vfill

\setcounter{footnote}{0}
\renewcommand{\thefootnote}{\arabic{footnote}}

\end{titlepage}

\tableofcontents

\section{introduction}
The WZNW model has been well studied through its relationship with other mathematical models\cite{Moore:1988qv,AlvarezGaume:1989aq,Gerasimov:1990fi}.
As a recent topic, there are the studies of the relations between the Racah matrices and the HOMFLY polynomials\cite{Smirnov:2009fs,Mironov:2011ym,Mironov:2011aa,Itoyama:2012qt,Itoyama:2012re,Anokhina:2013wka,Nawata:2013ppa}. 

Especially in the case of $sl_{2}$ WZNW model, as for its exchange relations, the coefficients have been computed explicitly for screened vertex operators possessing arbitrary weights. 
The coefficients are in accordance with the Racah-Wigner $q$-$6j$-symbol\cite{KiriResh1989,AlvarezGaume:1988vr,Itoyama:1989mw,Balog:1990dn,Itoh:1990bi,Itoh:1992sq,Smithies1995}.

On the other hand, in the case of $sl_{3}$ WZNW model, its exchange relations have not been studied very well.
When the rank of the group is two or more than two, the free field representation (or so-called Wakimoto representation\cite{Wakimoto:1986gf,KT1998}) possesses multiple number of bosonic fields.
And if we give a screened vertex operator including at least two kinds of screening charges, an ambiguity in ordering of its screening charges appears.

To begin with, one has to introduce six appropriate screened vertex operators, as shown in \eqref{eq:wznw_bases}, so that their exchange relation takes a closed form. 
Then the coefficients of the exchange relations coincide with the exchange relations of spider diagrams introduced in \cite{Archer:1992kc,1997q.alg....12003K}.
The spider diagram is a useful method to represent $U_{q}(sl_{3})$ in a graphical way. 
In this paper, we consider its semiclassical structure given by the limit where $q$ equals $1$.
We do not know any systematic way of handing this procedure in the case where $q$ is a root of unity.

The organization of the paper is as follows. 
In section \ref{ss:WZNWmodel}, first we introduce the $sl_{3}$ WZNW model in Wakimoto representation and define the vertex operators, the screening charges and the six appropriate screened vertex operators. 
In section \ref{ss:contour}, we demonstrate the contour integrals of the screening charges, and then find out the useful identities.
In section \ref{ss:rmatrix}, we list the results of calculations of exchange relations.
In section \ref{ss:spiders}, we summarize the spider diagram and its exchange relations.
In the appendix \ref{ss:calculationOfSpiderDiagrams}, we show concrete calculations of the spider diagram.

\section{WZNW model}\label{ss:WZNWmodel}
In this paper, we are concerned about the semiclassical limit of the exchange relation only.
The semiclassical limit implies the infinity limit in level $k$ of the $sl_{3}$ WZNW model.
Even within this simplification we still obtain a nontrivial exchange relation.

\subsection{definitions of free fields}
Latin letters take $1,2$, while greek letters take $1,2,3$.

It is well known that the affine Lie current of the $sl_{3}$ Lie algebra can be constructed by using the several free fields. 
As a introduction to WZNW model, there are references \cite{AlvarezGaume:1989aq,fuchs:book,yamadabook2006}  

We introduce free fields by the following formal Laurent series.
\begin{align}
	\phi_{i}(z)&=\frac{1}{\kappa}\left(q_{i}+(a_{i})_{0}\log z
		-\sum_{n\neq0}\frac{(a_{i})_{n}}{n}z^{-n}\right),\\
	a_{i}(z)
	&=\sum_{n\in\Integer}(a_{i})_{n}z^{-n-1},\\
	\beta_{\alpha}(z)&=\sum_{n\in\Integer}(\beta_{\alpha})_{n}z^{-n-1},\\
	\gamma_{\alpha}(z)&=\sum_{n\in\Integer}(\gamma_{\alpha})_{n}z^{-n}.
\end{align}
Here $\kappa=k+3$, and
 $q_{i}$, $(a_{i})_{n}$, $(\beta_{\alpha})$, and($\gamma_{\alpha})_{n}$ obey the following commutation relations.
\begin{align}
	[(a_{i})_{m},(a_{j})_{n}]&=\kappa m a_{ij}\delta_{m+n,0},\\
	[(a_{i})_{0},q_{j}]&=\kappa a_{ij},\\
	[(\beta_{\alpha})_{m},(\gamma_{\beta})_{n}]&=\delta_{\alpha\beta}\delta_{m+n,0}
\end{align}
$\delta_{m,n}$ is Kronecker delta,
and $a_{ij}$ is Cartan matrix of $sl_{3}$:
\begin{gather}\begin{split}
	a_{ij}=\begin{pmatrix}2&-1\\-1&2\end{pmatrix}.
\end{split}\end{gather}
Vacuum expectation value of an operator is written as
\begin{gather}\begin{split}
	\langle \mathcal{O}\rangle=\langle0|\mathcal{O}|0\rangle
\end{split}\end{gather}
The bracket is defined to satisfy
\begin{gather}\begin{split}
	\langle0|\{(a_{i})_{n<0},~
	            q_{i},~
	            (\beta_{\alpha})_{n<0},~
	            (\gamma_{\alpha})_{n\geq0}\}
	=0,~~~
	\{(a_{i})_{n\geq0},~
	 (\beta_{\alpha})_{n\geq0},~
	 (\gamma_{\alpha})_{n<0}\}|0\rangle=0,
	~~~\langle0|1|0\rangle=1.
\end{split}\end{gather}
The operator product expansions of the above free fields are 
\begin{align}
	&\phi_{i}(z)\phi_{j}(w)=\frac{a_{ij}}{\kappa}\log(z-w)+\cdots,\\
	&a_{i}(z)\phi_{j}(w)=\frac{a_{ij}}{z-w}+\cdots,\\
	&\phi_{i}(z)a_{j}(w)=-\frac{a_{ij}}{z-w}+\cdots,\\
	&\beta_{\alpha}(z)\gamma_{\beta}(w)=\frac{\delta_{\alpha\beta}}{z-w}+\cdots.
\end{align}
Here the terms omitted are the normal ordered terms which do not have poles at $z=w$.

Two vertex operators are introduced 
as an exponential function of the free fields $\phi_{i}$.
\begin{align}
	V_{1}(z)=:e^{\frac{1}{3}(2\phi_{1}(z)+\phi_{2}(z))}:~,\\
	V_{2}(z)=:e^{\frac{1}{3}(\phi_{1}(z)+2\phi_{2}(z))}:~.
\end{align}
By Wick theorem, they show singularity.
\begin{gather}\begin{split}
\label{eq:Wick}
	:e^{\phi_{i}(z)}::e^{\phi_{i}(w)}:
	=(z-w)^{a_{ij}/\kappa}:e^{\phi_{i}(z)}e^{\phi_{i}(w)}:
\end{split}\end{gather}

\subsection{definitions of bases and exchange relations}
Usually the screening charges are defined as following. 
\begin{align}
	\tilde{Q}_{1}(z)\equiv
	 \frac{1}{1-q^{-1}}
	 \oint_{\Gamma_{z}}
	  (\beta_{1}(w)
	   +\frac{1}{2}
	    \gamma_{2}(w)\beta_{3}(w))
	    e^{-\phi_{1}(w)}dw~,
\\
	\tilde{Q}_{2}(z)\equiv
	 \frac{1}{1-q^{-1}}
	 \oint_{\Gamma_{z}} 
	  (\beta_{2}(w)
	   -\frac{1}{2}
	    \gamma_{1}(w)\beta_{3}(w))
	    e^{-\phi_{2}(w)}dw~.
\end{align}
Here $q=e^{2\pi i/\kappa}$, and the integration contour $\Gamma_{z}$ starts with $z$, ends with $z$, and surrounds ordered vertex operators to its right.

To simplify the calculation of the contour integrals,
we use the screening charges which have no  $\beta_{\alpha}(z)$ and $\gamma_{\alpha}(z)$.
This simplification does not change the coefficients of exchange relations. 
\begin{gather}\begin{split}
\label{eq:screeningCharge}
	Q_{i}(z)\equiv
	 \frac{1}{1-q^{-1}}
	 \oint_{\Gamma_{z}}
	    s_{i}(z)dw~,
~~~
	s_{i}(z)\equiv e^{-\phi_{i}(z)}~.
\end{split}\end{gather}
While the factor $\frac{1}{1-q^{-1}}$ diverges to infinity when we take the classical limit ($q\to 1$), the screening charge $Q_{i}$ is finite thank to an infinitesimal factor comes from the contour integrals like as \eqref{eq:fundIntegral1}.

\begin{figure}[htbp]
\begin{center}
\includegraphics[scale=0.8]{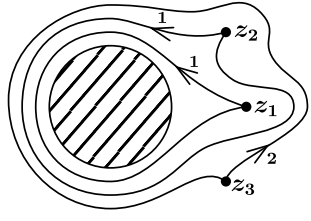}
\caption{{\bf contours}: The labels $1$ and $2$ near the arrows denote the subscripts of each screening charge. The central disk covers the vertex operators on the right hand of the three screening charges $Q_{2}(z_{3})Q_{1}(z_{2})Q_{1}(z_{1})$. }
\label{fig:contours}
\end{center}
\end{figure}

When we multiply by a number of screening charges
\begin{gather}\begin{split}
	Q_{2}(z_{3})Q_{1}(z_{2})Q_{1}(z_{1})\cdots,
\end{split}\end{gather}
these contours take the form as shown in Fig.\ref{fig:contours}. 

In the semiclassical limit $\kappa\to\infty$ ($q\to 1$),
 it is immaterial what branch we choose, because the factor $q^{r}$ ($r$ is a rational number) originating in branch becomes $1$ in this limit.

The following equations introduce the six screened vertex operators, whose exchange relations are take a closed form.
\begin{gather}
\boxed{
\begin{split}
\label{eq:wznw_bases}
	U^{(1,0)}(z)&\equiv V_{1}(z)~,~
\\
	U^{(-1,1)}(z)&\equiv V_{1}(z)Q_{1}(z)\frac{1}{J_{1}}~,
\\
	U^{(0,-1)}(z)
	&\equiv
	 V_{1}(z)\Big(
	     Q_{1}(z)Q_{2}(z)(J_{2}+1)
	    -Q_{2}(z)Q_{1}(z)J_{2}
	    \Big)
	    \frac{1}{J_{2}(J_{1}+J_{2}+1)}~,\\
	&= 
	V_{1}(z)Q_{12}(z)\frac{1}{J_{2}(J_{1}+J_{2}+1)}~,
\\
	U^{(0,1)}(z)&\equiv V_{2}(z)~,~
\\
	U^{(1,-1)}(z)&\equiv V_{2}(z)Q_{2}(z)\frac{1}{J_{2}}~,
\\
	U^{(-1,0)}(z)&\equiv
	 V_{1}(z)\Big(
	     Q_{2}(z)Q_{1}(z)(J_{1}+1)
	    -Q_{1}(z)Q_{2}(z)J_{1}
	    \Big)
	    \frac{1}{J_{1}(J_{1}+J_{2}+1)}~,\\
	&=
	V_{2}(z)Q_{21}(z)\frac{1}{J_{1}(J_{1}+J_{2}+1)}~.
\end{split}
}\end{gather}

Here the superscripts of $U$ are the weights of each operator respectively, and
\begin{gather}\begin{split}
	Q_{i\bar{i}}(z)\equiv
	\frac{1}{(1-q^{-1})^{2}}
	\oint_{\Gamma_{z}}dw_{1}s_{i}(w_{1})\oint_{\Gamma_{w_{1}}}dw_{2}s_{\bar{i}}(w_{2}),
\end{split}\end{gather}
where $i=1,2$, $\bar{1}\equiv2$ and $\bar{2}\equiv1$,
 and the operators $J_{i}\equiv(a_{i})_{0}$ count the number of the vertex operators to its right.
\begin{gather}\begin{split}
	&[J_{i},V_{j}(z)]=\delta_{ij}V_{1}(z)~,\\
	&[J_{i},U^{\lambda}(z)]=\lambda_{i}U^{\lambda}(z)~,\\
\end{split}\end{gather}
Here and below, the indices $\lambda_{i}~(i= 1,2,3,4)$ run over the weights $\{(1,0),~(-1,1),~(0,-1),~(0,1),~(1,-1),~(-1,0)\}$.

We define $|\mu\rangle$ which is an eigenvector of $J_{i}$.
\begin{gather}\begin{split}
	J_{1}|\mu\rangle=m|\mu\rangle~,~~~	
	J_{2}|\mu\rangle=n|\mu\rangle~.	
\end{split}\end{gather}
here $\mu=(m,n)$ and
\begin{gather}\begin{split}
	|\mu\rangle
	\equiv
	\prod_{i=1}^{m}V_{1}(x_{i})
	\prod_{j=1}^{n}V_{2}(y_{j})|0\rangle~,	
\end{split}\end{gather}
We consider $x_{i}$ and $y_{j}$ as arbitrary positions.

The exchange relations between screened vertex operators \eqref{eq:wznw_bases} are expressed as
\begin{gather}\begin{split}
\label{eq:exchange_relations}
	U^{\lambda_{1}}(z)U^{\lambda_{2}}(w)|\mu\rangle
	=\sum_{\lambda_{3}\lambda_{4}}~
	[c_{\textrm{WZNW}}(\mu)]^{\lambda_{1}\lambda_{2}}_{\lambda_{3}\lambda_{4}}~
	U^{\lambda_{3}}(w)U^{\lambda_{4}}(z)|\mu\rangle.
\end{split}\end{gather} 
In the section \ref{ss:rmatrix}, we will list its concrete expression.


\subsection{contour integrals}
\label{ss:contour}

In this section, we show some identities of the contour integrals carried out in the semiclassical limit $\kappa\to \infty$.
As a reference for the contour integrals, we have used \cite{AoKi:1994}. 

In this section, we confirm the \emph{identities} listed in \eqref{eq:Qide:S} to \eqref{eq:tetraId}.
These identities are used to calculate the exchange relations.
Commutativity:
\begin{gather}
\label{eq:Qide:S}
	[Q_{i}(z),V_{\bar{i}}(z_{1})]=0~,
\\
	[Q_{i}(z),V_{\bar{i}}(z_{1},z_{2})]=0~,
\\
	[Q_{i}(z),I(z_{1},z_{2})]=0~,
\\
	[Q_{i}(z),I_{j}(z_{1},z_{2},z_{3})]=0~,
\end{gather}
Symmetry:
\begin{gather}
	V_{i}(z_{1},z_{2})=-V_{i}(z_{2},z_{1}),~
\\
	I(z_{1},z_{2})=I(z_{2},z_{1}),~
\\
	I_{i}(z_{1},z_{2},z_{3})=I_{i}(z_{2},z_{3},z_{1})
	=-I_{i}(z_{2},z_{1},z_{3})~,
\end{gather}
\begin{gather}\begin{split}
\label{eq:tetraId}
	I_{i}(z_{1},z_{2},z_{3})V_{\bar{i}}(z_{4})+
	I(z_{1},z_{4})V_{\bar{i}}(z_{2},z_{3})+
	I(z_{2},z_{4})V_{\bar{i}}(z_{3},z_{1})+
	I(z_{3},z_{4})V_{\bar{i}}(z_{1},z_{2})
	&=0~,
\end{split}\end{gather}
where $i,j=1,2$, $\bar{1}\equiv2,~\bar{2}\equiv1$ and
\begin{gather}\begin{split}
	V_{\bar{i}}(z_{1},z_{2})
	&:=
	[V_{i}(z_{1})Q_{i}(z_{1}),V_{i}(z_{2})]~,
\\
	I(z_{1},z_{2})
	&:=
	[V_{1}(z_{1})Q_{1}(z_{1}),~[Q_{2}(z_{1}),~V_{2}(z_{2})]]~,
\\
	I_{i}(z_{1},z_{2},z_{3})
	&:=
	[V_{i}(z_{1})Q_{i}(z_{1}),~[Q_{\bar{i}}(z_{1}),~
		[V_{i}(z_{2})Q_{i}(z_{2}),~V_{i}(z_{3})]]]~.
\end{split}\end{gather}

To express the contour integrals by using a graphical way,
we list the rules of the diagrams as followings.
\begin{itemize}
\item The points denote the positions of vertex operators.
\item The labels $1$ and $2$ near the points, denote the indices of the vertex operators.
\item The lines denote the contours.
\item The arrows denote the directions of integrals.
\item The labels $1$ and $2$ near the arrows, denote the indices of the screening charges.
\end{itemize}

From now on, let us carry out the calculation for the above identities.

The first commutation relation as below often appears in the actual calculations of the exchange relations.
\begin{gather}\begin{split}
\label{eq:VQV}
	V_{2}(z,0)
	=
	[V_{1}(z)Q_{1}(z),~ V_{1}(0)]
	=
	\left[ V_{1}(z)
	\left(\frac{1}{1-q^{-1}}\oint_{\Gamma_{z}}dw~s_{1}(w)\right),~V_{1}(0)\right].\\
\end{split}\end{gather}
Since the $\kappa$ is large enough to converge the integral around the pole, one can take the contour in accordance with Cauchy's integral theorem as following.
\begin{align}
\label{eq:fundIntegral1}
\includegraphics[scale=0.8]{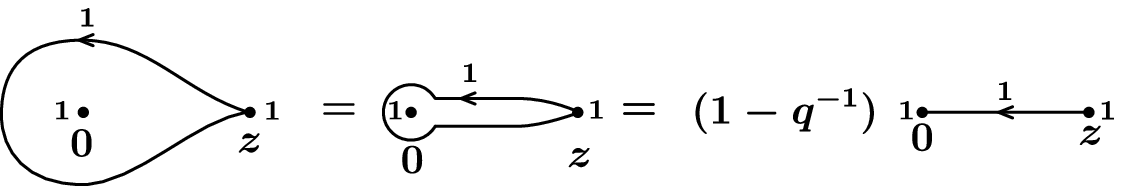}
\end{align}
While the factor $1-q^{-1}$ is infinitesimal, there is factor $1/(1-q^{-1})$ in the definitions of screening charges \eqref{eq:screeningCharge}, so that the contour integral \eqref{eq:VQV} becomes finite.

Next, we consider the following commutation relation.
\begin{gather}\begin{split}
\label{eq:QVQV}
	[Q_{1}(z_{2}),V_{2}(z,0)]=0~,
\end{split}\end{gather}
The contour of $V_{2}(z,0)$ is similar to \eqref{eq:fundIntegral1}. 
By adding the contour coming from $Q_{1}(z_{2})$, we obtain following diagram. 
\begin{gather}\begin{split}
\label{eq:qvqv}
\includegraphics[scale=0.8]{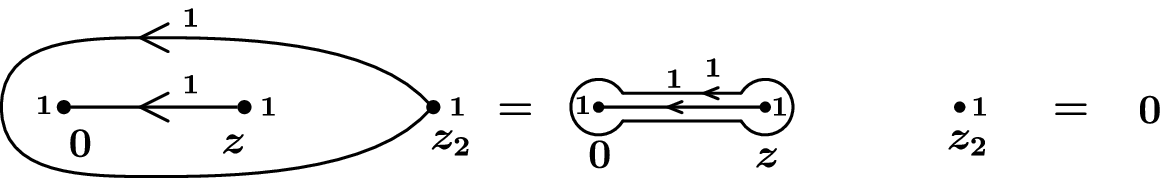}
\\
\end{split}\end{gather}
The second equality is shown by cutting off the contour into four contours as following.
\begin{gather}\begin{split}
\label{eq:path_cancellation}
\includegraphics[scale=0.8]{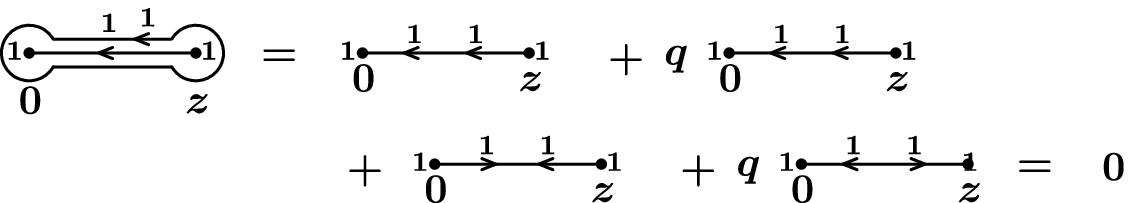}
\\
\end{split}\end{gather}
Hence the equation \eqref{eq:QVQV} is given. 
We take no account of the overall phase coming from the determination of the branch.
The contour having two arrows, means a double line integral respect to two variable as keeping these ordering.

Next useful contour integral is composed of two kinds of vertex operators $V_{1}$ and $V_{2}$, and two kinds of screening charges $Q_{1}$ and $Q_{2}$ as below.
\begin{gather}\begin{split}
\label{eq:V1Q1Q2V2}
	I(z,0)=
	[V_{1}(z)Q_{1}(z),[Q_{2}(z),V_{2}(0)]]
\end{split}\end{gather}
The contour lines for $Q_{1}$ and $Q_{2}$ are deformed to one line having two arrows:
\begin{align}
\label{eq:vqqv}
\includegraphics[scale=0.6]{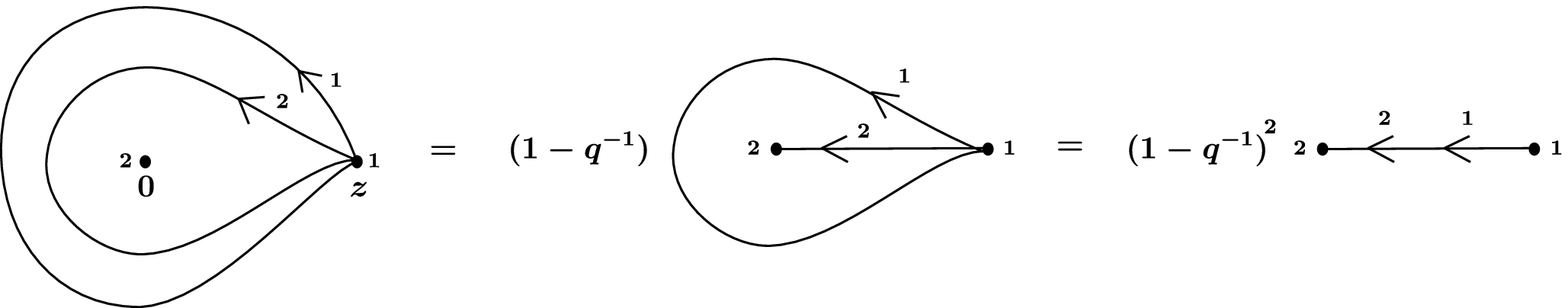}
\end{align}
$I(z,0)$ commutes with both screening charges $Q_{1}$ and $Q_{2}$.
\begin{gather}\begin{split}
	[Q_{i}(z_{1}),I(z,0)]=0~,
\end{split}\end{gather}
This commutatively is concerned by a procedure similar to  \eqref{eq:path_cancellation}:
\begin{align}
\label{eq:qvqqv}
\includegraphics[scale=0.6]{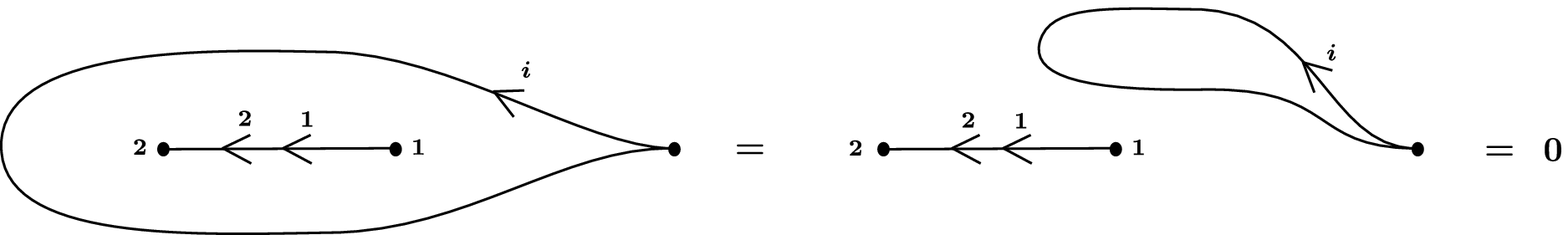}
\end{align}

There are another type of integrals, which also commutes with both screening charges $Q_{1}(z)$ and $Q_{2}(z)$.
\begin{gather}\begin{split}
	[Q_{i}(z),I_{1}(z_{1},z_{2},z_{3})]=0.
\end{split}\end{gather}
This commutatively is concerned by a procedure similar to  \eqref{eq:path_cancellation} too.

The contour of $I_{1}(z_{1},z_{2},z_{3})$ can be drawn as below.  
\begin{align}
\label{eq:I1}
\includegraphics[scale=0.6]{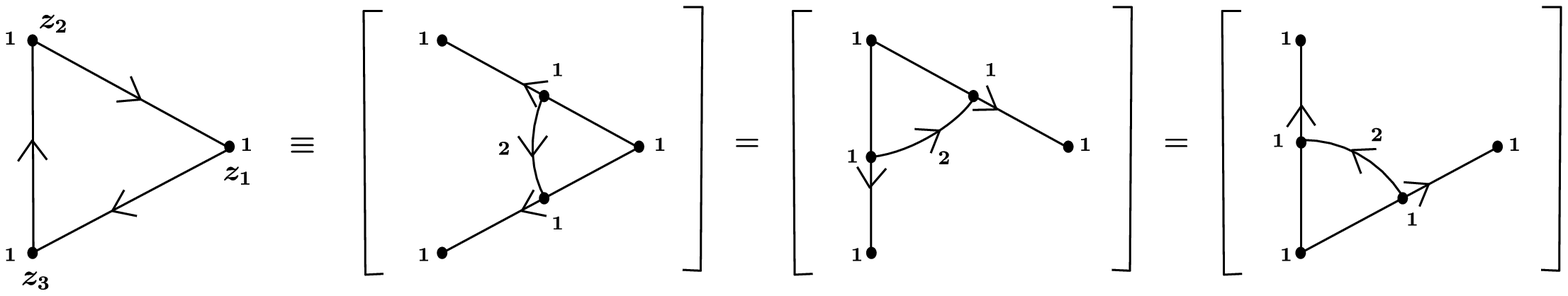},
\end{align}
In these diagrams, the contour of $Q_{2}$ goes from the position of one screen current $s_{1}$ to the position of another screen current $s_{1}$.  
By mutually replacing the labels $1$ and $2$ in \eqref{eq:I1}, we obtain $I_{2}(z_{1},z_{2},z_{3})$.

The identity \eqref{eq:tetraId} can be shown by applying  Cauchy's integral theorem to the following contours corresponding to \eqref{eq:tetraId}.
\begin{align}
\label{eq:tetra:id}
\includegraphics[scale=0.6]{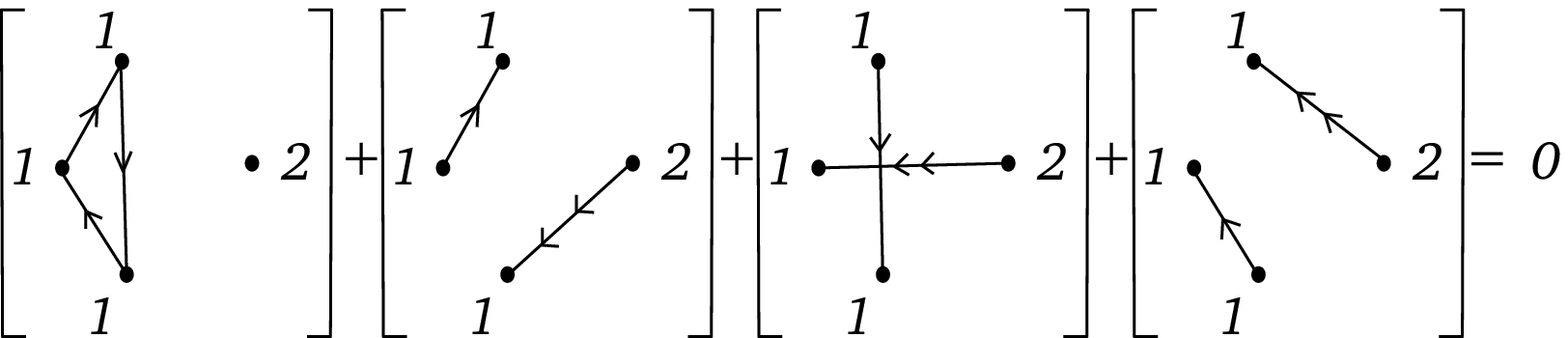}~,
\end{align}
while we have omitted the labels near the arrows in \eqref{eq:tetra:id},  we assume that the labels $1,2$ can be assigned according to following rules.
\begin{gather}\begin{split}
\includegraphics[scale=0.6]{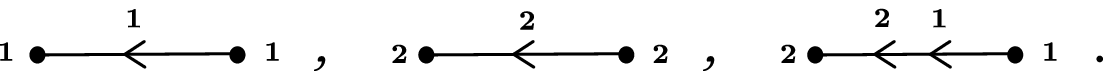}
\end{split}\end{gather}

\subsection{exchange relations}\label{ss:rmatrix}
In this section, we list the concrete expressions of exchange relations between the six screened vertex operators \eqref{eq:wznw_bases}.
These calculations are performed by comparing the contours in both side. Then we use the identities introduced in previous subsection.

\subsubsection*{exchange relations between $\{U^{(1,0)},~U^{(-1,1)},~U^{(0,-1)}\}$}
\begin{gather}\begin{split}
\label{eq:exchange_relations_UU}
	U^{(1,0)}(z)U^{(1,0)}(w)|\mu\rangle
	 &=U^{(1,0)}(w)U^{(1,0)}(z)|\mu\rangle,\\
	U^{(1,0)}(z)U^{(-1,1)}(w)|\mu\rangle
	 &=U^{(-1,1)}(w)U^{(1,0)}(z)|\mu\rangle
	   +
	   \frac{1}{m+1}\langle \psi|U^{(1,0)}(w)U^{(-1,1)}(z)|\mu\rangle\\
	U^{(-1,1)}(z)U^{(1,0)}(w)|\mu\rangle
	 &=\frac{m(m+2)}{(m+1)^{2}}
	   U^{(1,0)}(w)U^{(-1,1)}(z)|\mu\rangle
	   -
	   \frac{1}{m+1}
	   U^{(-1,1)}(w)U^{(1,0)}(z)|\mu\rangle\\
	U^{(-1,1)}(z)U^{(-1,1)}(w)|\mu\rangle
	 &=U^{(-1,1)}(w)U^{(-1,1)}(z)|\mu\rangle\\
	U^{(1,0)}(z)U^{(0,-1)}(w)|\mu\rangle
	 &=U^{(0,-1)}(w)U^{(1,0)}(z)|\mu\rangle
	   +\frac{1}{m+n+2}
	    U^{(1,0)}(w)U^{(0,-1)}(z)|\mu\rangle\\
	U^{(0,-1)}(z)U^{(1,0)}(w)|\mu\rangle
	 &=\frac{(m+n+1)(m+n+3)}{(m+n+2)^{2}}
	   U^{(1,0)}(w)U^{(0,-1)}(z)|\mu\rangle
	 \\
	 &~~~~
	  -\frac{1}{m+n+2}U^{(0,-1)}(w)U^{(1,0)}(z)|\mu\rangle\\
	U^{(0,-1)}(z)U^{(-1,1)}(w)|\mu\rangle
	 &=U^{(-1,1)}(w)U^{(0,-1)}(z)|\mu\rangle
	 -\frac{1}{n+1}U^{(0,-1)}(w)U^{(-1,1)}(z)|\mu\rangle\\\
	U^{(-1,1)}(z)U^{(0,-1)}(w)|\mu\rangle
	 &=\frac{n(n+2)}{(n+1)^{2}}
	   U^{(0,-1)}(w)U^{(-1,1)}(z)|\mu\rangle
	 +\frac{1}{n+1}
	   U^{(-1,1)}(w)U^{(0,-1)}(z)|\mu\rangle\\
	U^{(0,-1)}(z)U^{(0,-1)}(w)|\mu\rangle
	 &=U^{(0,-1)}(w)U^{(0,-1)}(z)|\mu\rangle
\end{split}\end{gather}

\subsubsection*{exchange relations between $\{U^{(0,1)},U^{(1,-1)},U^{(-1,0)}\}$}

\begin{gather}\begin{split}
\label{eq:exchange_relations_UbUb}
	U^{(0,1)}(z)U^{(0,1)}(w)|\mu\rangle
	 &=U^{(0,1)}(w)U^{(0,1)}(z)|\mu\rangle,\\
	U^{(0,1)}(z)U^{(1,-1)}(w)|\mu\rangle
	 &=U^{(1,-1)}(w)U^{(0,1)}(z)|\mu\rangle
	   +
	   \frac{1}{n+1}
	   U^{(0,1)}(w)U^{(1,-1)}(z)|\mu\rangle\\
	U^{(1,-1)}(z)U^{(0,1)}(w)|\mu\rangle
	 &=\frac{n(n+2)}{(n+1)^{2}}
	 U^{(0,1)}(w)U^{(1,-1)}(z)|\mu\rangle   
	   -
	 \frac{1}{n+1}
	 U^{(1,-1)}(w)U^{(0,1)}(z)|\mu\rangle\\
	U^{(1,-1)}(z)U^{(1,-1)}(w)|\mu\rangle
	 &=U^{(1,-1)}(w)U^{(1,-1)}(z)|\mu\rangle\\
	U^{(0,1)}(z)U^{(-1,0)}(w)|\mu\rangle
	 &=
	 U^{(-1,0)}(w)U^{(0,1)}(z)|\mu\rangle
	 +\frac{1}{m+n+2}
	 U^{(1,0)}(w)U^{(-1,0)}(z)|\mu\rangle\\
	U^{(-1,0)}(z)U^{(0,1)}(w)|\mu\rangle
	 &=\frac{(m+n+1)(m+n+3)}{(m+n+2)^{2}}
	 U^{(0,1)}(w)U^{(-1,0)}(z)|\mu\rangle
	 \\
	 &~~~~
	 -\frac{1}{m+n+2}
	 U^{(-1,0)}(w)U^{(0,1)}(z)|\mu\rangle\\
	U^{(-1,0)}(z)U^{(1,-1)}(w)|\mu\rangle
	 &=U^{(1,-1)}(w)U^{(-1,0)}(z)|\mu\rangle
	 -\frac{1}{m+1}
	 U^{(-1,0)}(w)U^{(1,-1)}(z)|\mu\rangle\\
	U^{(1,-1)}(z)U^{(-1,0)}(w)|\mu\rangle
	 &=\frac{m(m+2)}{(m+1)^{2}}
	 U^{(-1,0)}(w)U^{(1,-1)}(z)|\mu\rangle  
	 +\frac{1}{m+1}
	 U^{(1,-1)}(w)U^{(-1,0)}(z)|\mu\rangle\\
	U^{(-1,0)}(z)U^{(-1,0)}(w)|\mu\rangle
	 &=U^{(-1,0)}(w)U^{(-1,0)}(z)|\mu\rangle
\end{split}\end{gather}
Up to now we have given the exchange relations  between $\{U^{(1,0)},U^{(-1,1)},U^{(0,-1)}\}$ and ones between $\{U^{(0,1)},U^{(1,-1)},U^{(-1,0)}\}$ with closed form, respectively. 

\subsubsection*{exchange relations between $\{U^{(1,0)},U^{(-1,1)},U^{(0,-1)}\}$ and $\{U^{(0,1)},U^{(1,-1)},U^{(-1,0)}\}$}

\begin{gather}\begin{split}
\label{eq:exchange_relations_UUb}
	U^{(0,1)}(z)U^{(1,0)}(w)|\mu\rangle
	&=
	U^{(1,0)}(w)U^{(0,1)}(z)|\mu\rangle,
\\
	U^{(1,-1)}(z)U^{(1,0)}(w)|\mu\rangle
	&=U^{(1,0)}(w)U^{(1,-1)}(z)|\mu\rangle,
\\
	U^{(0,1)}(z)U^{(-1,1)}(w)|\mu\rangle
	&=U^{(-1,1)}(w)U^{(0,1)}(z)|\mu\rangle,
\\
	U^{(-1,0)}(z)U^{(1,0)}(w)|\mu\rangle
	&=\frac{m(m+2)(m+n+1)(m+n+3)}{(m+1)^{2}(m+n+2)^{2}}
	   U^{(1,0)}(w)U^{(-1,0)}(z)|\mu\rangle
	\\&~~~~~
	+
	\frac{m+2}{(m+1)(m+n+2)}
	 U^{(0,-1)}(w)U^{(0,1)}(z)|\mu\rangle,
	\\&~~~~~
	-
	 \frac{n(m+n+3)}{(m+1)(n+1)(m+n+2)}
	 U^{(-1,1)}(w)U^{(1,-1)}(z)|\mu\rangle
\\
	U^{(1,-1)}(z)U^{(-1,1)}(w)|\mu\rangle
	&=
	  \frac{n}{n+1}
	U^{(-1,1)}(w)U^{(1,-1)}(z)|\mu\rangle
	-
	  \frac{1}{n+1}
	U^{(0,-1)}(w)U^{(0,1)}(z)|\mu\rangle
	\\&~~~~~
	+
	  \frac{(n+2)(m+n+1)}
	       {(m+1)(n+1)(m+n+2)},
	U^{(1,0)}(w)U^{(-1,0)}(z)|\mu\rangle
\\
	U^{(0,1)}(z)U^{(0,-1)}(w)|\mu\rangle
	&=
	U^{(0,-1)}(w)U^{(0,1)}(z)|\mu\rangle
	+
	\frac{1}{n+1}
	U^{(-1,1)}(w)U^{(1,-1)}(z)|\mu\rangle
	\\&~~~~~
	-
	\frac{m}{(m+1)(m+n+2)}
	U^{(1,0)}(w)U^{(-1,0)}(z)|\mu\rangle
\\
	U^{(1,-1)}(z)U^{(0,-1)}(w)|\mu\rangle
	&=	U^{(0,-1)}(w)U^{(1,-1)}(z)|\mu\rangle
\\
	U^{(0,-1)}(z)U^{(1,-1)}(w)|\mu\rangle
	&=U^{(1,-1)}(w)U^{(0,-1)}(z)|\mu\rangle
\\
	U^{(0,-1)}(z)U^{(-1,0)}(w)|\mu\rangle
	&=U^{(-1,0)}(w)U^{(0,-1)}(z)|\mu\rangle.
\\
\end{split}\end{gather}

So we have demonstrated that the exchange relations between the six bases\eqref{eq:wznw_bases}, can be expressible in closed form

\section{spider diagrams}\label{ss:spiders}
In this section, we give a brief review of the spider diagrams considered in \cite{Archer:1992kc,1997q.alg....12003K}.  
We take a limit $q\to 1$ to obtain semiclassical limit.
In this case, the fundamental relations for the spider diagrams are defined as following.
\begin{align}
\label{eq:spider:fund}
\includegraphics[scale=0.8]{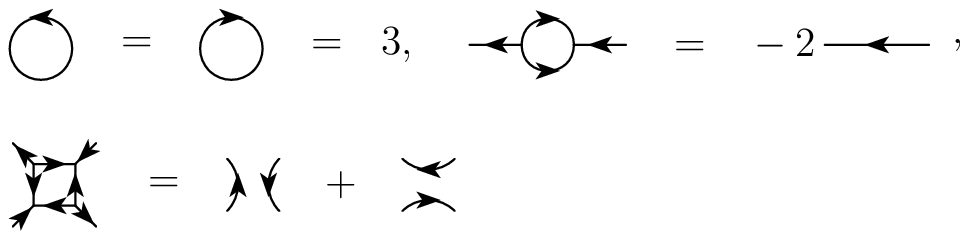}
\end{align}
It is defined that the crossing lines can be decomposed into two diagrams.
\begin{align}
\includegraphics[scale=0.8]{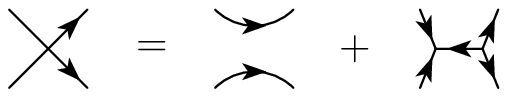}
\end{align}
The lines crossing twice come untied.
\begin{align}
\label{eq:spider_cross_square}
\includegraphics[scale=0.8]{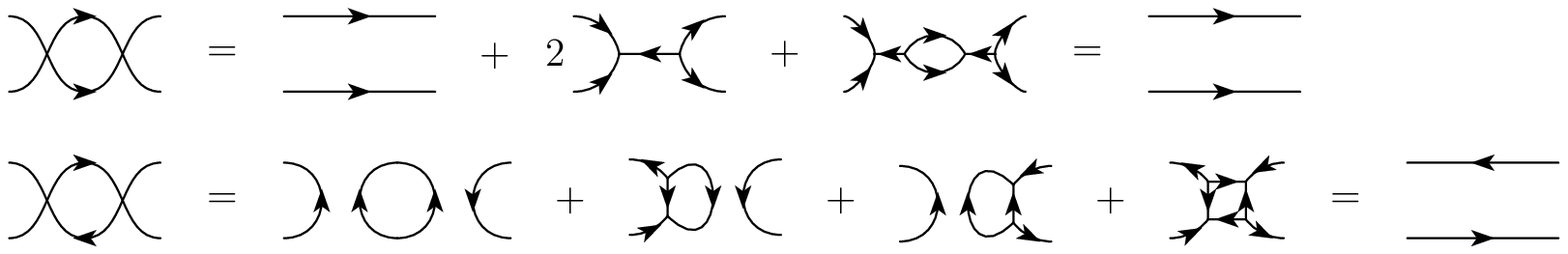}
\end{align}
There are diagrams, called \emph{clasp}, which possess property of the projection operator as following.
\begin{align}
\label{eq:projection2rr}
\includegraphics[scale=0.8]{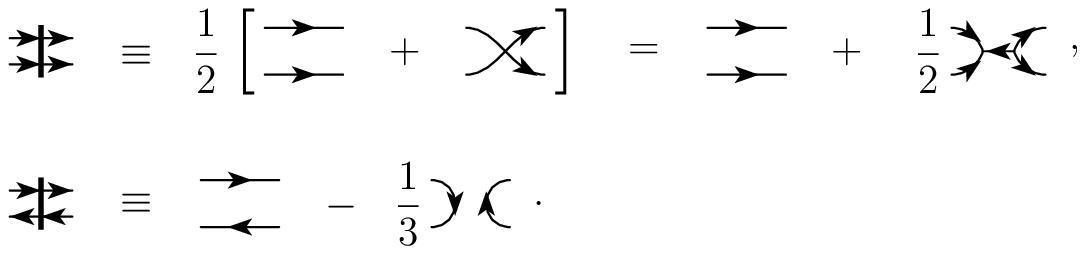}
\end{align}
\begin{align}
\includegraphics[scale=0.8]{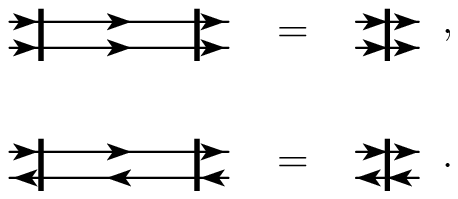}
\end{align}
In general, one can introduce the clasps having much more external lines.
For simplification, we represent the internal clasps as below. 
\begin{align}
\label{eq:projection_n}
\includegraphics[scale=0.8]{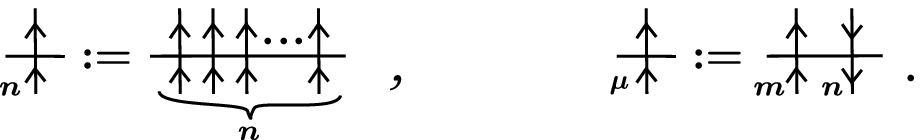}
\end{align}
where $\mu=(m,n)$. 
In this paper, the internal clasp satisfies the equations.
\begin{align}
\label{eq:projection_id}
\eqfig{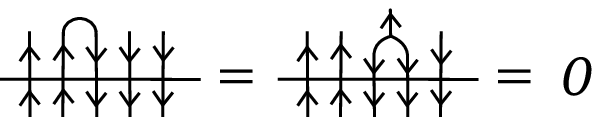}~~.
\end{align}
The clasps are obtained by the mathematical inductions.
\begin{align}
\label{eq:projection_n_ind}
\includegraphics[scale=0.8]{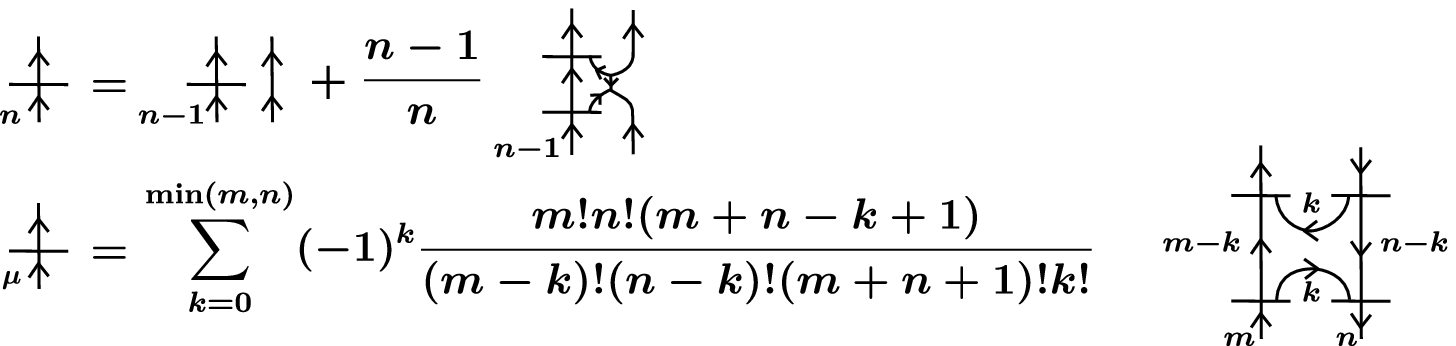}
\end{align}

In this paper, we use only first two terms of the expansion \eqref{eq:projection_n_ind}, because of that we consider only simple spider diagrams like as \eqref{eq:spider_bases}.
\begin{align}
\label{eq:projection_n_ind:2}
\eqfig{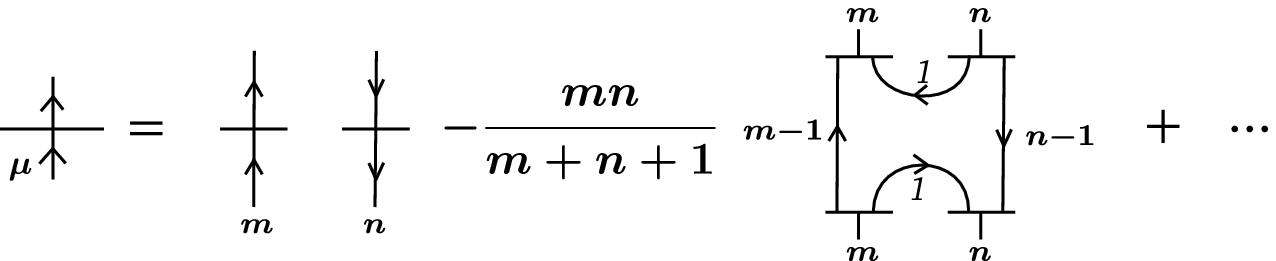}
\end{align}
\subsection{exchange relations}
\label{ss:exchange:spider}
We introduce the six spider diagrams like as
\begin{gather}\begin{split}
\label{eq:spider_bases}
&\raisebox{-25pt}{\includegraphics[scale=0.8]{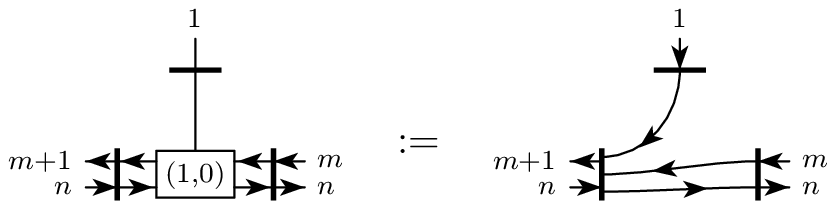}}~,~~~
\raisebox{-25pt}{\includegraphics[scale=0.8]{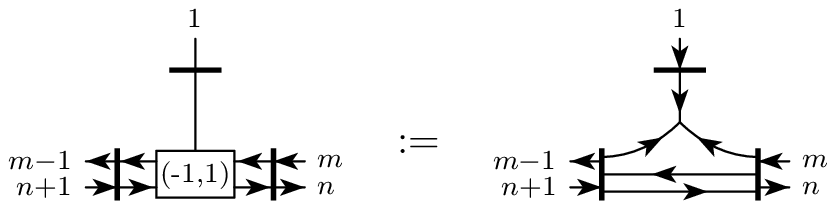}}~,\\
&\raisebox{-25pt}{\includegraphics[scale=0.8]{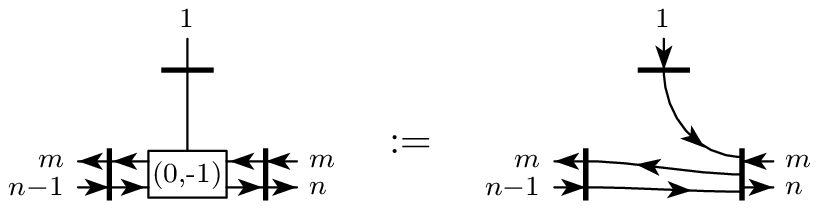}}~,~~~
\raisebox{-25pt}{\includegraphics[scale=0.8]{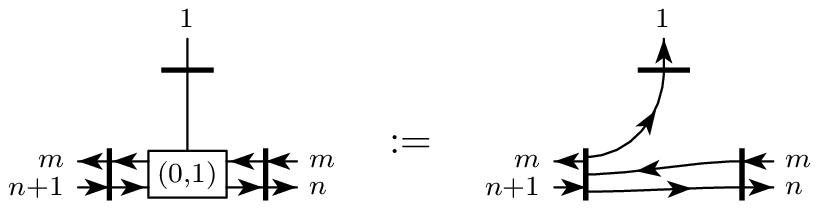}}~,\\
&\raisebox{-25pt}{\includegraphics[scale=0.8]{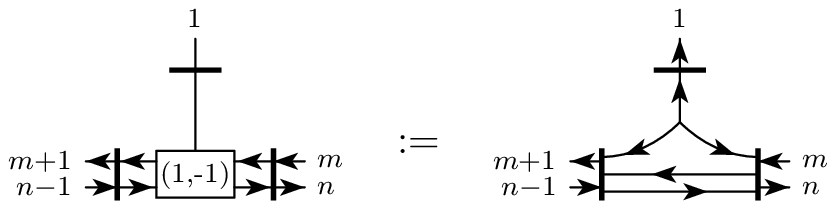}}~,~~~
\raisebox{-25pt}{\includegraphics[scale=0.8]{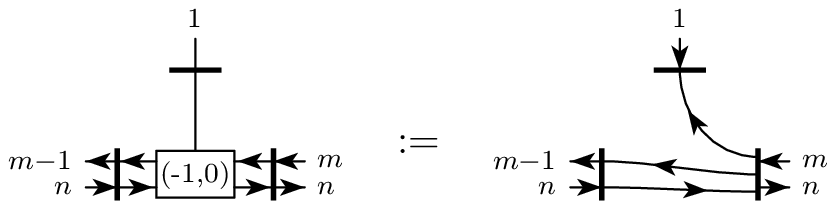}}~,
\end{split}\end{gather}
where we refer $\lambda\in\{(1,0),(-1,1),(0,-1),(0,1),(1,-1),(-1,0)\}$ as weights.  
The spider diagrams possessing a weight $\lambda$ changes the weight $\mu=(m,n)$ to $\mu+\lambda$.

Now we define the exchange relations for spider diagrams by introducing the following equation.
\begin{gather}\begin{split}
\label{eq:spider_6j-symbols}
\raisebox{-25pt}{\includegraphics[scale=0.8]{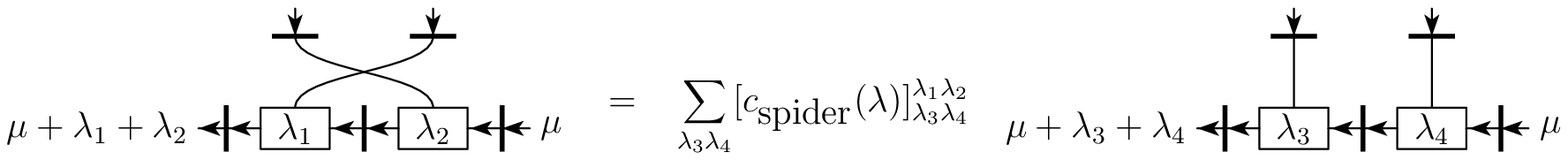}}~,
\end{split}\end{gather}
where $\mu=(m,n)$, $\lambda_{i}=\{(1,0),(-1,1),(0,-1),(0,1),(1,-1),(-1,0)\}$.
The left side of equation\eqref{eq:spider_6j-symbols} is a twisting spider diagram.
On the other hand, the right side is an untwisting spider diagram.   
The comparing both side is fulfilled by expanding the central clasps according to \eqref{eq:projection_n_ind}. 

Next we list the exchange relations of the spider diagrams.
These equations come from the intermediate expressions in appendix \ref{ss:calculationOfSpiderDiagrams}. 

The following equations are the exchange relations between spider diagrams\eqref{eq:spider_bases} having the weights $(1,0)$, $(-1,1)$ and $(0,-1)$. 
\begin{align}
\label{eq:spider_U1U1_twist_untwist}
&\eqfig{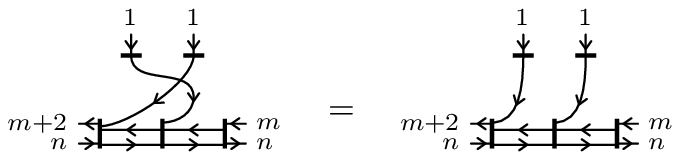}\\
\label{eq:spider_U1U2_twist_untwist}
&\eqfig{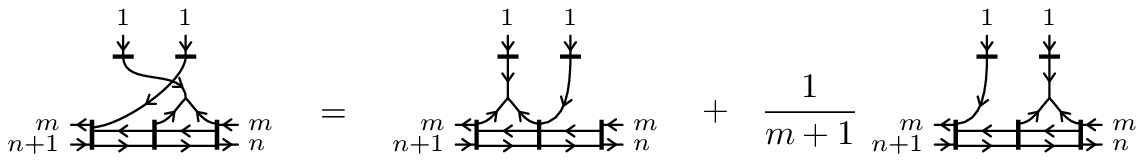}\\
\label{eq:spider_U2U1_twist_untwist}
&\eqfig{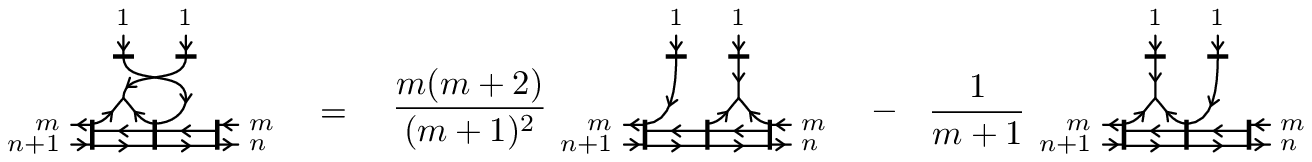}\\
\label{eq:spider_U2U2_twist_untwist}
&\eqfig{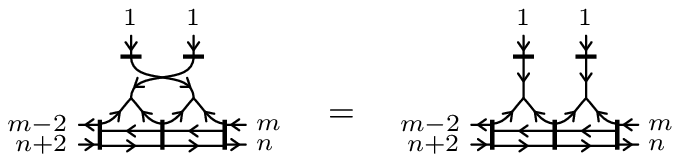}\\
\label{eq:spider_U1U3_twist_untwist}
&\eqfig{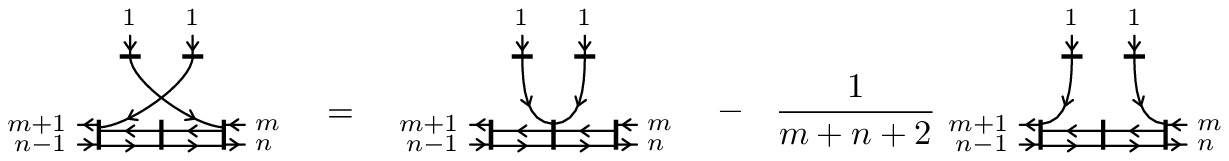}\\
\label{eq:spider_U3U1_twist_untwist}
&\eqfig{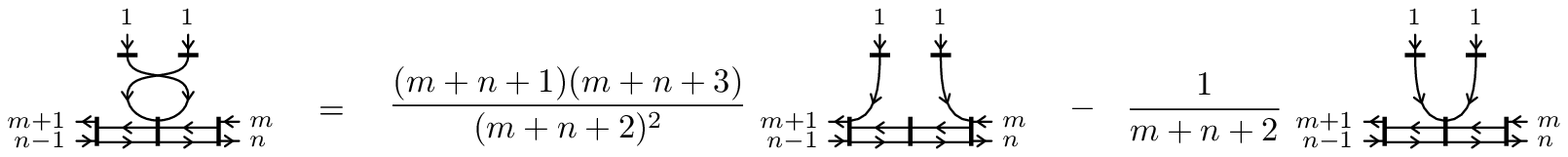}\\
\label{eq:spider_U2U3_twist_untwist}
&\eqfig{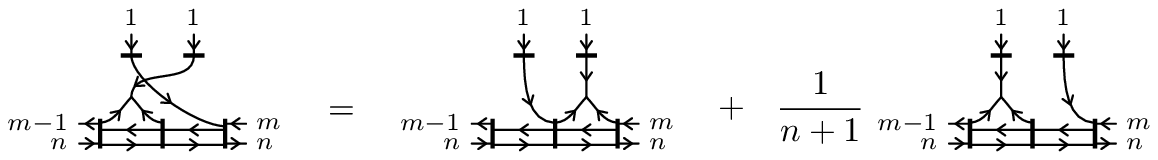}\\
\label{eq:spider_U3U2_twist_untwist}
&\eqfig{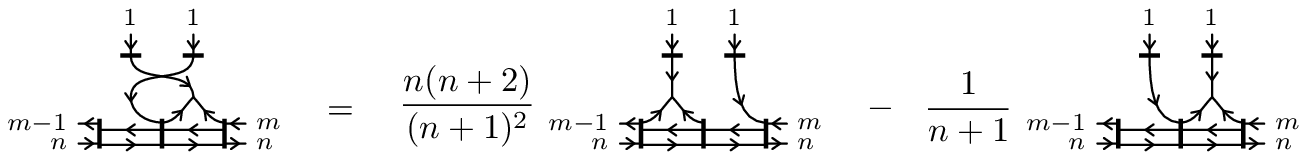}\\
\label{eq:spider_U3U3_twist_untwist}
&\eqfig{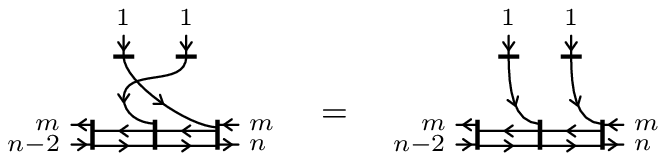}
\end{align}

The above coefficients correspond to the ones in \eqref{eq:exchange_relations_UU}. 

The following equations are the exchange relations between spider diagrams\eqref{eq:spider_bases} having the weights $(0,1)$, $(1,-1)$ and $(-1,0)$. 
\begin{align}
\label{eq:spider_Ub1Ub1_twist_untwist}
&\eqfig{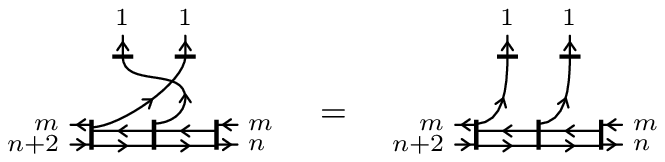}\\
\label{eq:spider_Ub1Ub2_twist_untwist}
&\eqfig{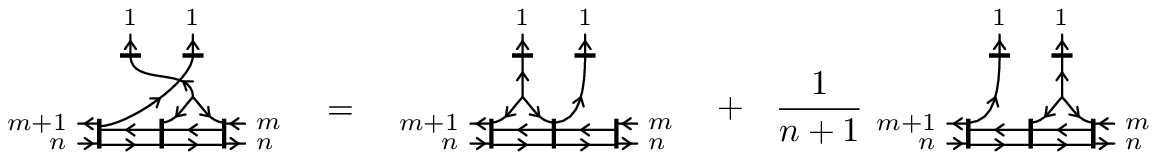}\\
\label{eq:spider_Ub2Ub1_twist_untwist}
&\eqfig{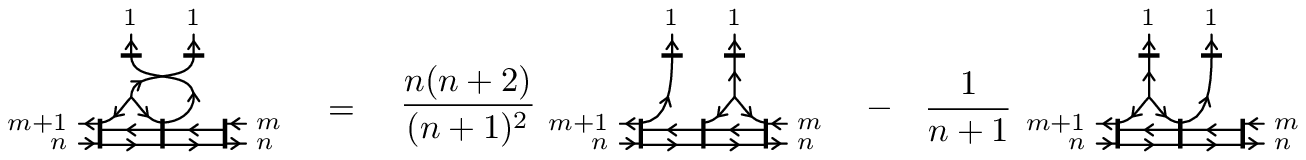}\\
\label{eq:spider_Ub2Ub2_twist_untwist}
&\eqfig{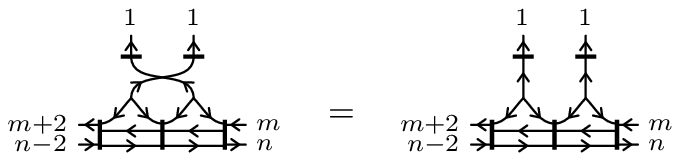}\\
\label{eq:spider_Ub1Ub3_twist_untwist}
&\eqfig{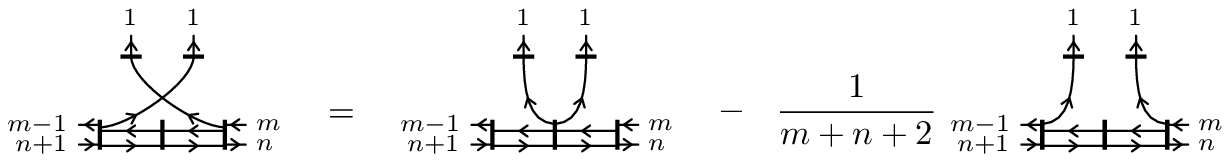}\\
\label{eq:spider_Ub3Ub1_twist_untwist}
&\eqfig{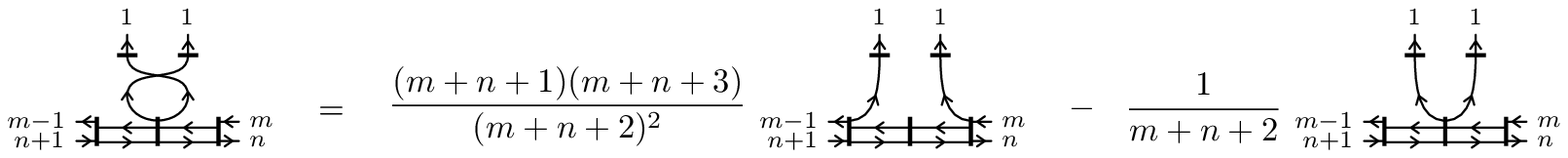}\\
\label{eq:spider_Ub2Ub3_twist_untwist}
&\eqfig{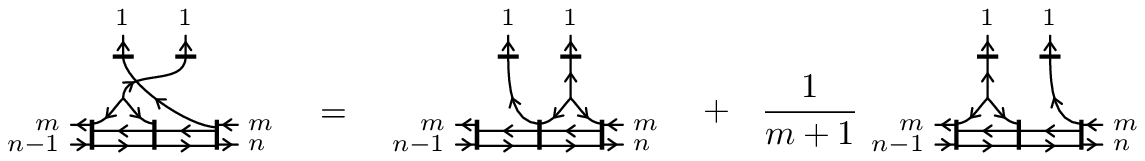}\\
\label{eq:spider_Ub3Ub2_twist_untwist}
&\eqfig{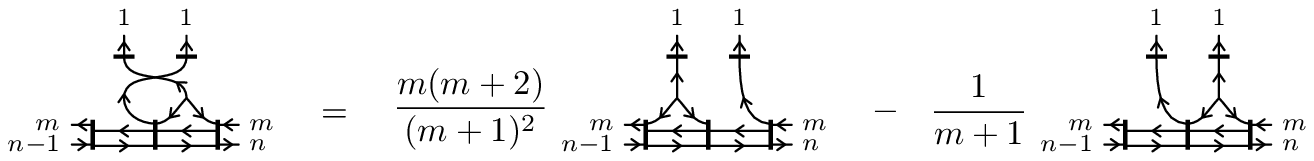}\\
\label{eq:spider_Ub3Ub3_twist_untwist}
&\eqfig{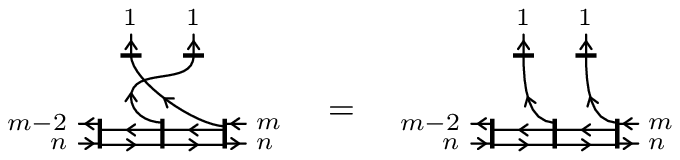}
\end{align}
The above coefficients correspond to the ones in \eqref{eq:exchange_relations_UbUb}.

The following equations are the exchange relations between spider diagrams\eqref{eq:spider_bases} having the weights 
$\{(1,0)$, $(-1,1)$,$(0,-1)\}$
and $\{(0,1)$, $(1,-1)$,$(-1,0)\}$.
 \begin{align}
\label{eq:spider_Ub1U1_twist_untwist}
&\eqfig{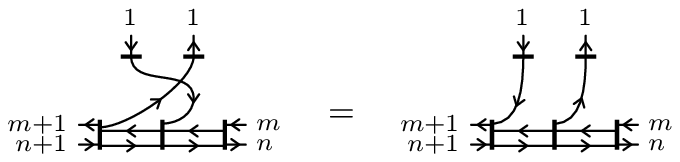}\\
\label{eq:spider_Ub2U1_twist_untwist}
&\eqfig{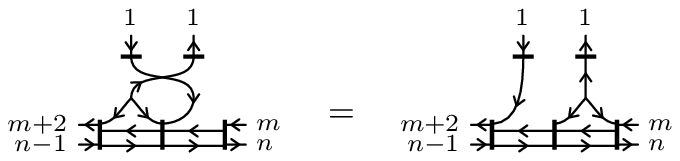}\\
\label{eq:spider_Ub1U2_twist_untwist}
&\eqfig{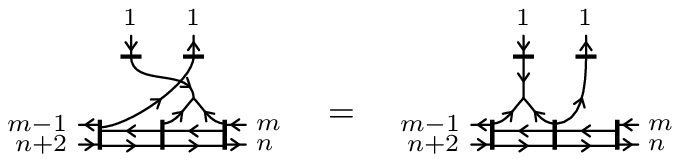}\\
\label{eq:spider_Ub3U1_twist_untwist}
&\eqfig{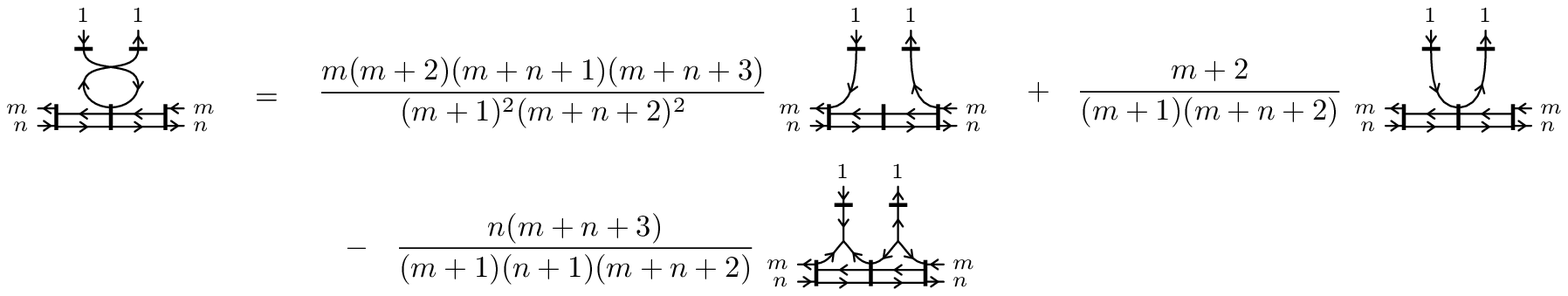}\\
\label{eq:spider_Ub1U3_twist_untwist}
&\eqfig{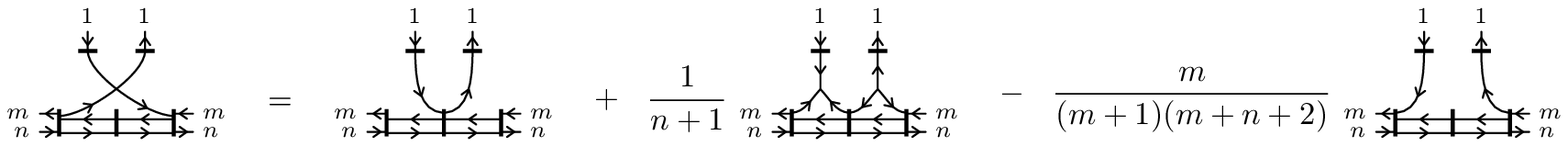}\\
\label{eq:spider_Ub2U2_twist_untwist}
&\eqfig{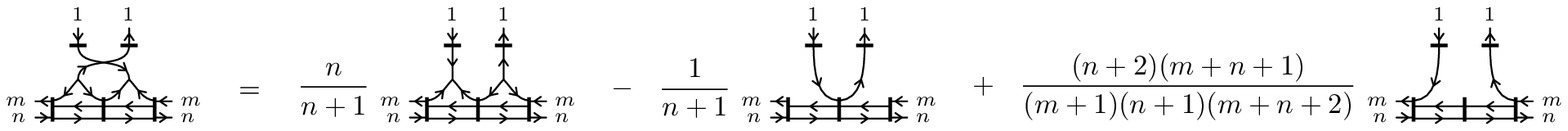}\\
\label{eq:spider_Ub3U2_twist_untwist}
&\eqfig{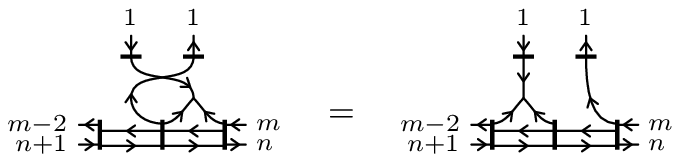}\\
\label{eq:spider_Ub2U3_twist_untwist}
&\eqfig{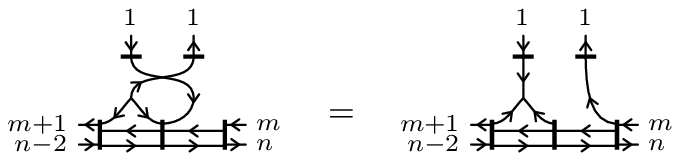}\\
\label{eq:spider_Ub3U3_twist_untwist}
&\eqfig{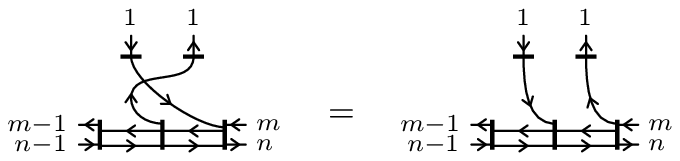}\\
\end{align}
The above coefficients correspond to the ones in \eqref{eq:exchange_relations_UUb}.

\section{correspondence}
\label{ss:correspondence}
In the previous two sections, we have calculated the exchange relations in both the $sl_{3}$ WZNW model and the spider diagram sides in semiclassical limit.
As a result, there is the correspondence between the two coefficients as following form. 
\begin{gather}\begin{split}
\label{eq:correspondences}
[c_{\textrm{WZNW}}(\mu)]^{\lambda_{1}\lambda_{2}}_{\lambda_{3}\lambda_{4}}
	=
[c_{\textrm{spider}}(\mu)]^{\lambda_{1}\lambda_{2}}_{\lambda_{3}\lambda_{4}}~.
\end{split}\end{gather}
These coefficients are defined in \eqref{eq:exchange_relations} and \eqref{eq:spider_6j-symbols}.
In appendix \ref{ss:calculationOfSpiderDiagrams}, we give
 the intermediate expressions of spider diagrams.

\section*{acknowledgment}
The author acknowledges helpful discussions with Hiroshi Itoyama as well as continual encouragement and careful reading of the manuscript. He also thanks Takeshi Oota  for helpful comments. He is supported by the scholarship of Graduate School of Science in Osaka City University for doctoral students.

\begin{appendix}

\section{calculations of spider diagrams}
\label{ss:calculationOfSpiderDiagrams}
In this appendix, we simplify both twisted spider diagrams and untwisted ones to compare each other.
The simplification is performed by expanding the central clasps according to \eqref{eq:projection_n_ind:2} and by using the fundamental equations \eqref{eq:spider:fund}.

All the twisted spider diagrams can be expressible in terms of untwisted ones finally. 
These results are listed in Sec.\ref{ss:exchange:spider}.

\subsection*{weight $(m,n)\to (m+2,n)$}
The twisted diagram is
\begin{align}
\label{eq:spider_U1U1_twist}
&\eqfig{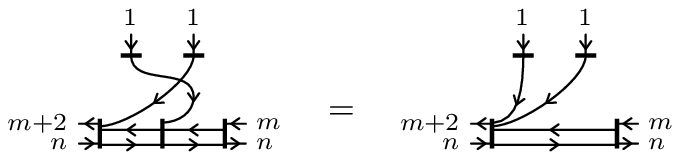}
\end{align}
The untwisted diagram is
\begin{align}
&\label{eq:spider_U1U1}
\eqfig{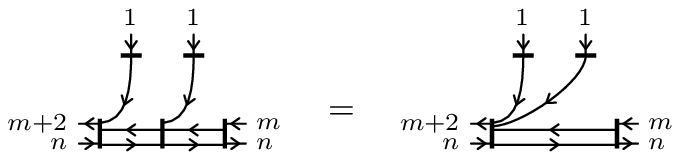}
\end{align}

\subsection*{weight $(m,n)\to (m,n+1)$}
The twisted diagrams are
\begin{align}
&\label{eq:spider_U1U2_twist}
\eqfig{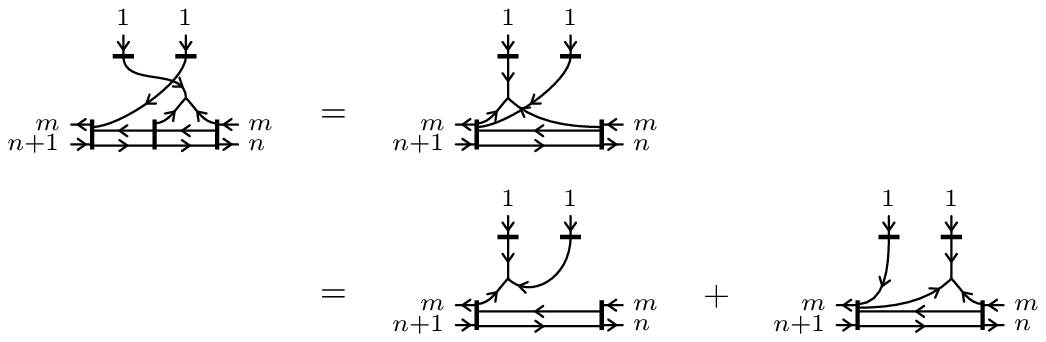}\\
&\label{eq:spider_U2U1_twist}
\eqfig{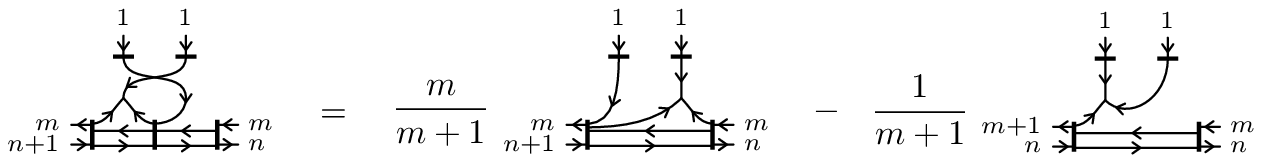}
\end{align}
The untwisted diagrams are
\begin{align}
&\label{eq:spider_U1U2}
\eqfig{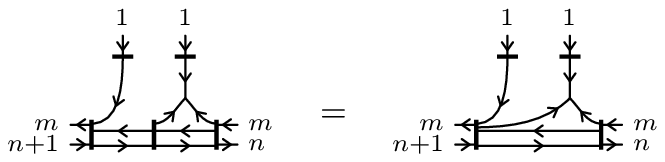}\\
&\label{eq:spider_U2U1}
\eqfig{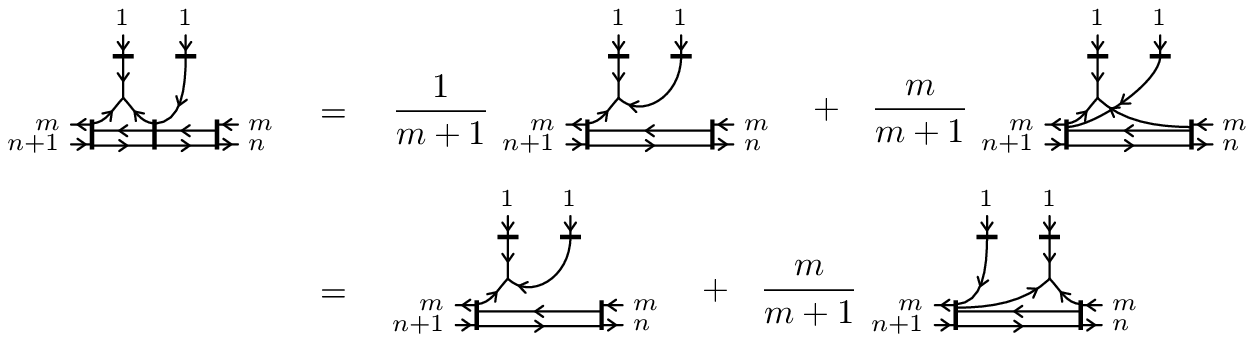}
\end{align}

\subsection*{weight $(m,n)\to (m+1,n-1)$}
The twisted diagrams are
\begin{align}
&\label{eq:spider_U3U1_twist}
\eqfig{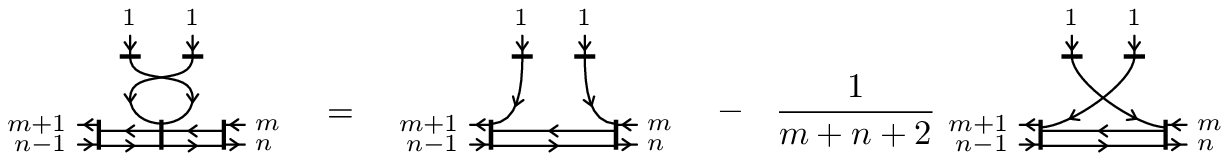}\\
&\label{eq:spider_U1U3_twist}
\eqfig{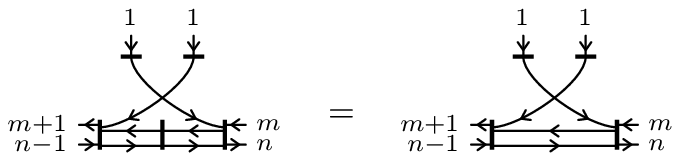}
\end{align}
The untwisted diagrams are
\begin{align}
&\label{eq:spider_U3U1}
\eqfig{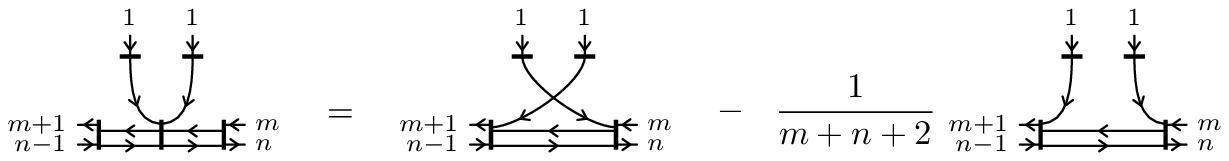}\\
&\label{eq:spider_U1U3}
\eqfig{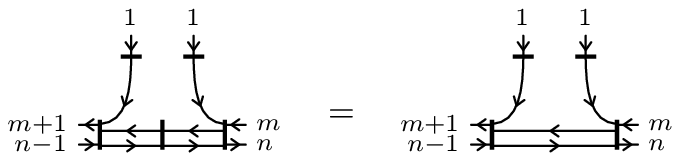}
\end{align}

\subsection*{weight $(m,n)\to (m-1,n)$}
The twisted diagrams are
\begin{align}
&\label{eq:spider_U2U3_twist}
\eqfig{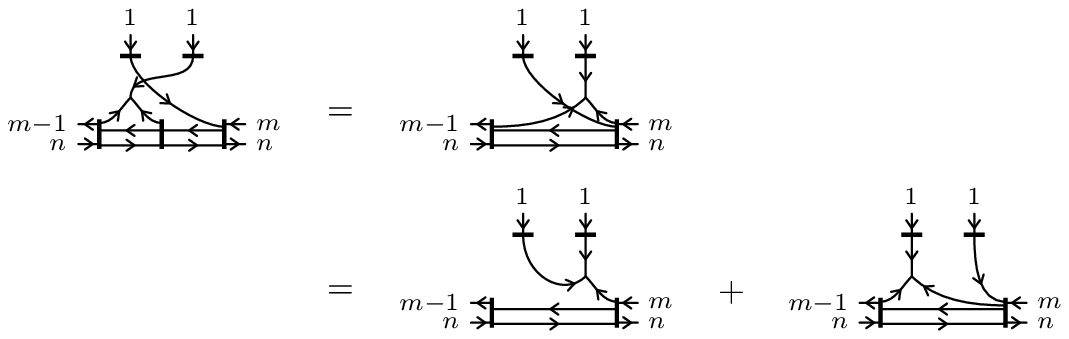}\\
&\label{eq:spider_U3U2_twist}
\eqfig{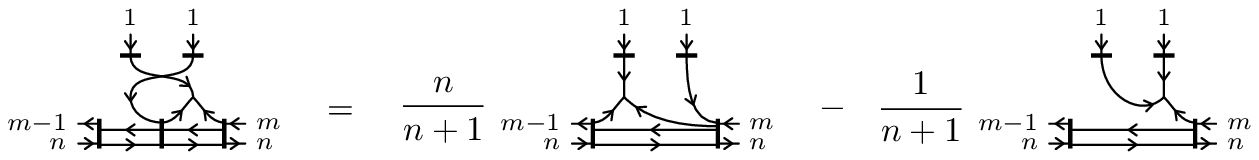}
\end{align}
The untwisted diagrams are
\begin{align}
&\label{eq:spider_U3U2}
\eqfig{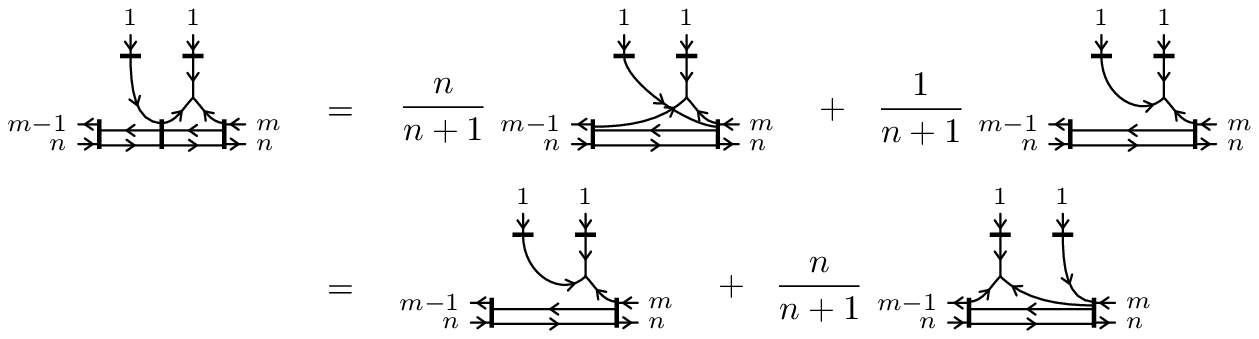}\\
&\label{eq:spider_U2U3}
\eqfig{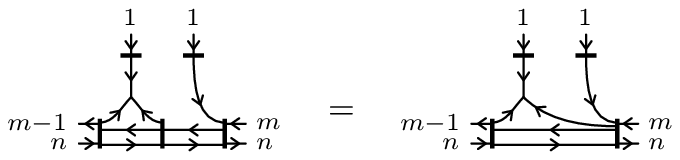}
\end{align}

\subsection*{weight $(m,n)\to (m,n-2)$}
The twisted diagram is
\begin{align}
&\label{eq:spider_U3U3_twist}
\eqfig{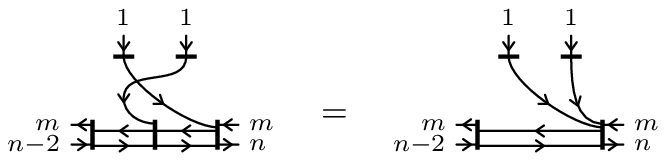}
\end{align}
The untwisted diagram is
\begin{align}
&\label{eq:spider_U3U3}
\eqfig{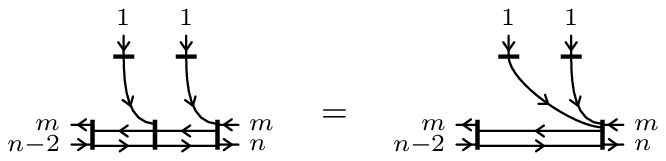}
\end{align}

\subsection*{weight $(m,n)\to (m-2,n+2)$}
The twisted diagram is
\begin{align}
&\label{eq:spider_U2U2}
\eqfig{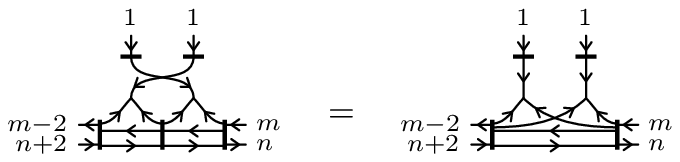}
\end{align}
The untwisted diagram is
\begin{align}
&\label{eq:spider_U2U2}
\eqfig{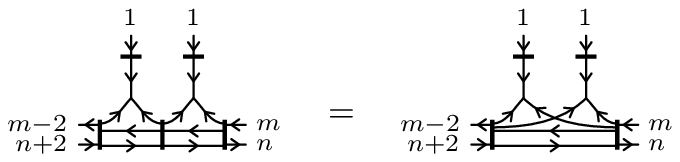}
\end{align}


\subsection*{weight $(m,n)\to (m,n+2)$}
The twisted diagram is
\begin{align}
\label{eq:spider_Ub1Ub1_twist}
&\eqfig{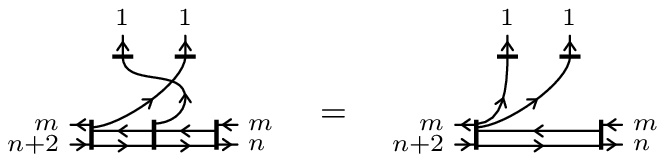}
\end{align}
The untwisted diagram is
\begin{align}
&\label{eq:spider_Ub1Ub1}
\eqfig{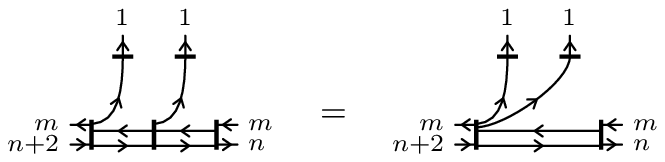}
\end{align}

\subsection*{weight $(m,n)\to (m+1,n)$}
The twisted diagrams are
\begin{align}
&\label{eq:spider_Ub1Ub2_twist}
\eqfig{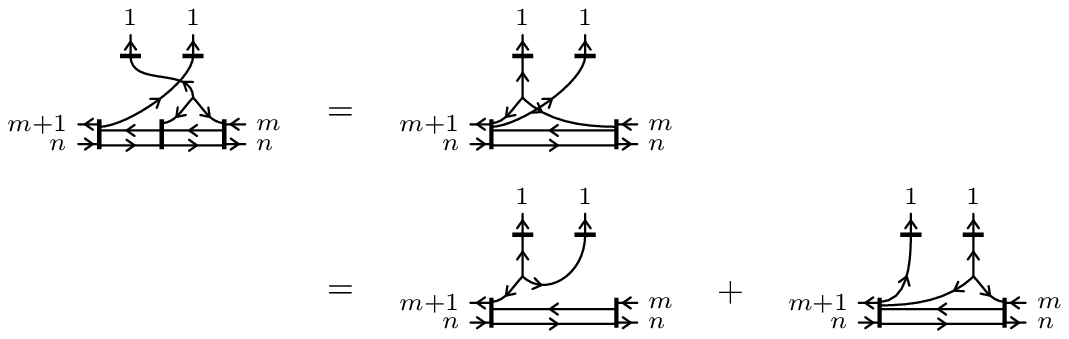}\\
&\label{eq:spider_Ub2Ub1_twist}
\eqfig{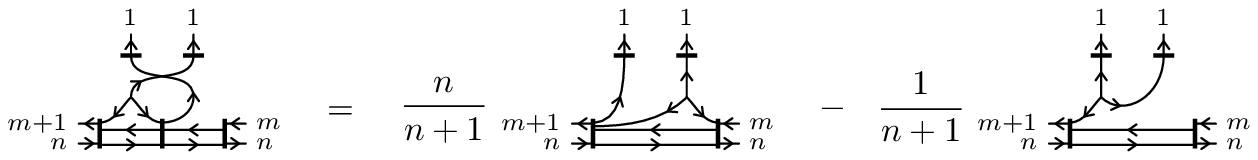}
\end{align}
The untwisted diagrams are
\begin{align}
&\label{eq:spider_Ub1Ub2}
\eqfig{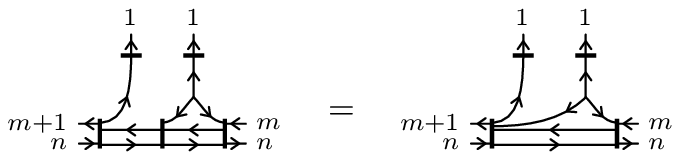}\\
&\label{eq:spider_Ub2Ub1}
\eqfig{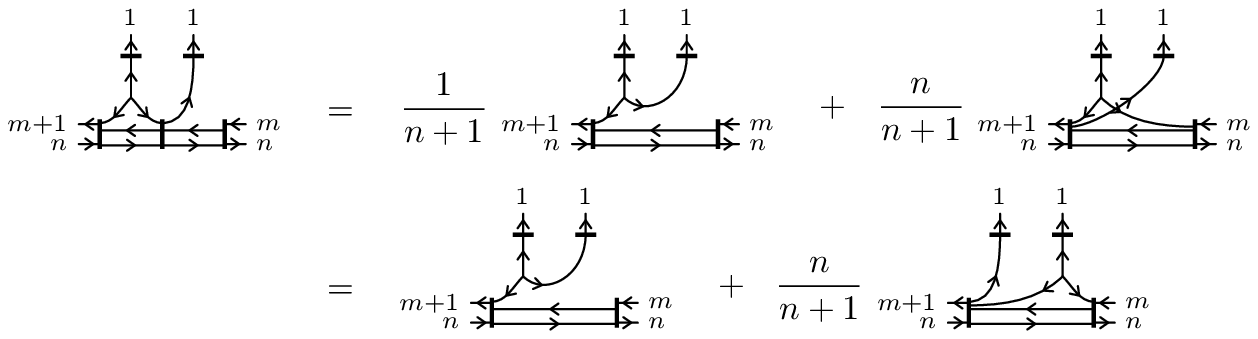}
\end{align}

\subsection*{weight $(m,n)\to (m+2,n-2)$}
The twisted diagram is
\begin{align}
&\label{eq:spider_Ub2Ub2}
\eqfig{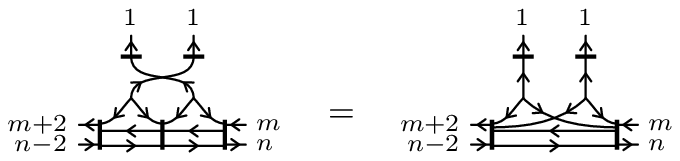}
\end{align}
The untwisted diagram is
\begin{align}
&\label{eq:spider_Ub2Ub2}
\eqfig{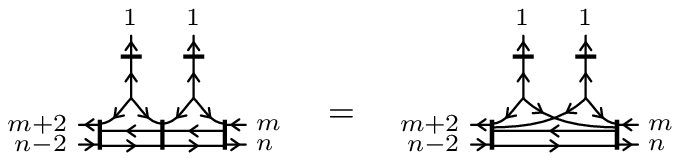}
\end{align}

\subsection*{weight $(m,n)\to (m-1,n+1)$}
The twisted diagrams are
\begin{align}
&\label{eq:spider_Ub3Ub1_twist}
\eqfig{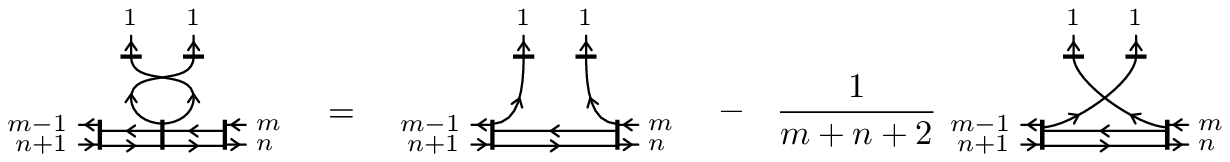}\\
&\label{eq:spider_Ub1Ub3_twist}
\eqfig{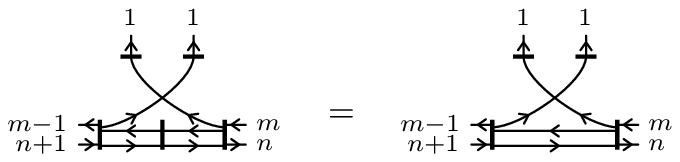}
\end{align}
The untwisted diagrams are
\begin{align}
&\label{eq:spider_Ub3Ub1}
\eqfig{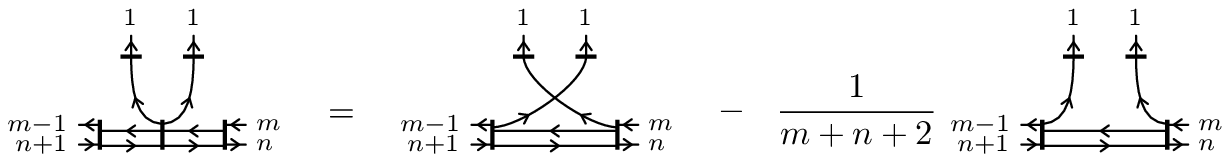}\\
&\label{eq:spider_Ub1Ub3}
\eqfig{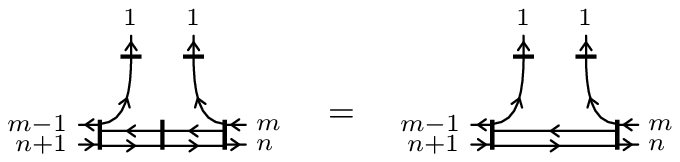}
\end{align}

\subsection*{weight $(m,n)\to (m,n-1)$}
The twisted diagrams are
\begin{align}
&\label{eq:spider_Ub2Ub3_twist}
\eqfig{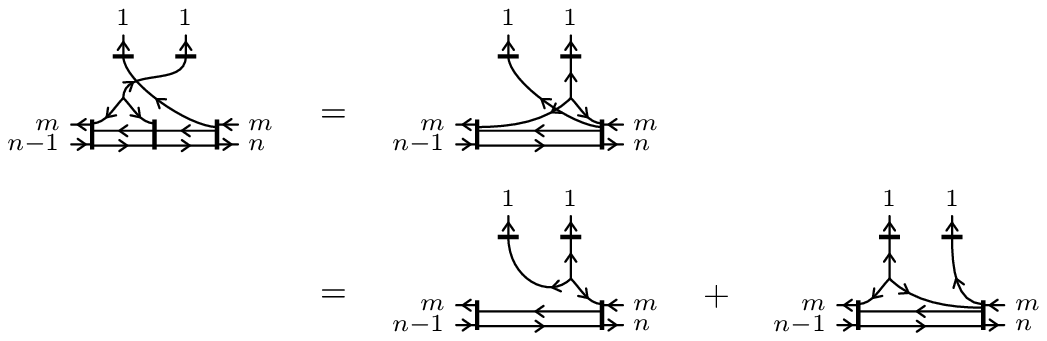}\\
&\label{eq:spider_Ub3Ub2_twist}
\eqfig{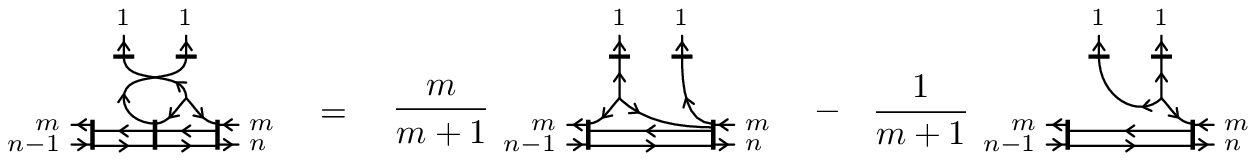}
\end{align}
The untwisted diagrams are
\begin{align}
&\label{eq:spider_Ub3Ub2}
\eqfig{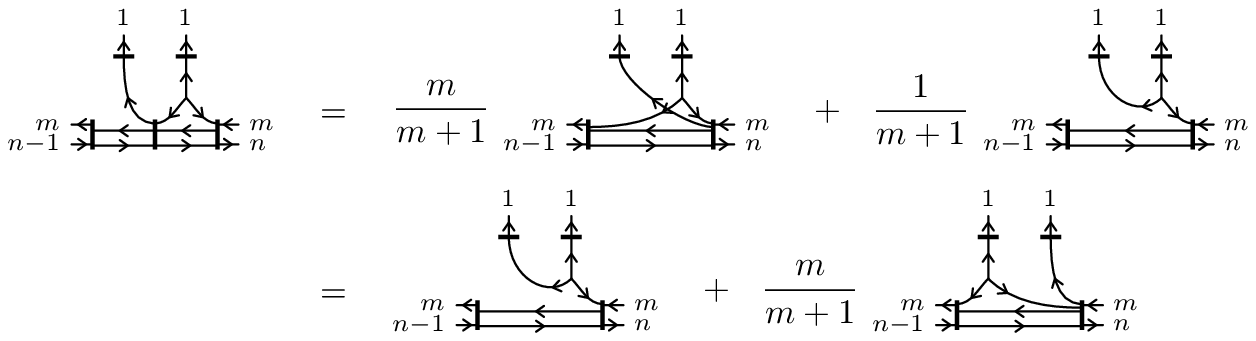}\\
&\label{eq:spider_Ub2Ub3}
\eqfig{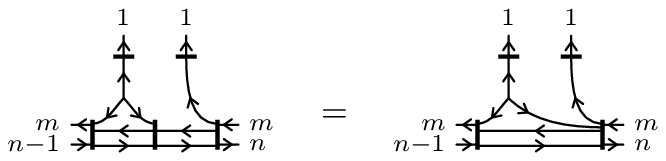}
\end{align}

\subsection*{weight $(m,n)\to (m-2,n)$}
The twisted diagram is
\begin{align}
&\label{eq:spider_Ub3Ub3_twist}
\eqfig{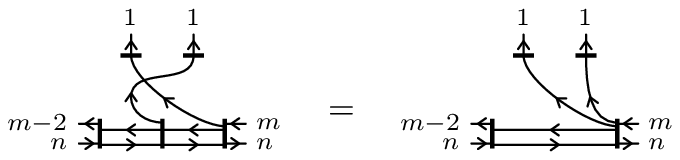}
\end{align}
The untwisted diagram is
\begin{align}
&\label{eq:spider_Ub3Ub3}
\eqfig{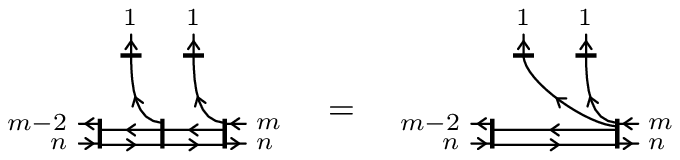}
\end{align}

\subsection*{weight $(m,n)\to (m+1,n+1)$}
The twisted diagram is
\begin{align}
\label{eq:spider_Ub1U1_twist}
&\eqfig{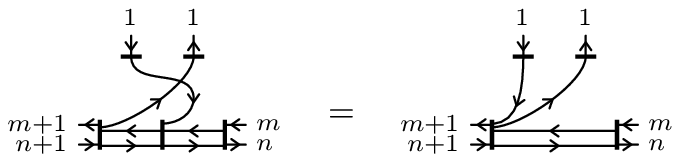}
\end{align}
The untwisted diagram is
\begin{align}
&\label{eq:spider_U1Ub1}
\eqfig{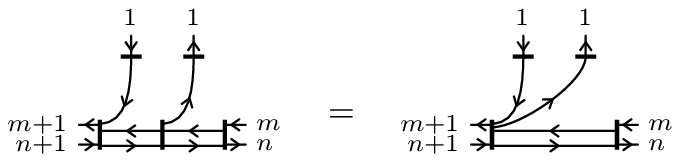}
\end{align}

\subsection*{weight $(m,n)\to (m+2,n-1)$}
The twisted diagram is
\begin{align}
\label{eq:spider_Ub2U1_twist}
&\eqfig{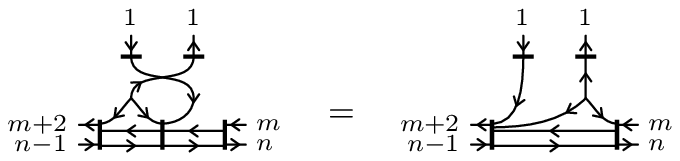}
\end{align}
The untwisted diagram is
\begin{align}
&\label{eq:spider_U1Ub2}
\eqfig{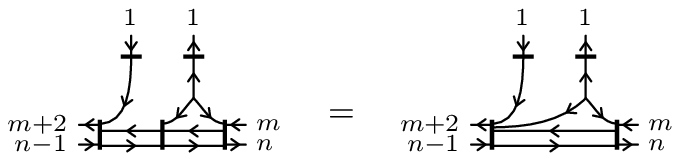}
\end{align}

\subsection*{weight $(m,n)\to (m-1,n+2)$}
The twisted diagram is
\begin{align}
\label{eq:spider_Ub1U2_twist}
&\eqfig{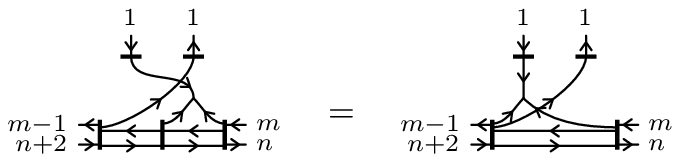}
\end{align}
The untwisted diagram is
\begin{align}
&\label{eq:spider_U2Ub1}
\eqfig{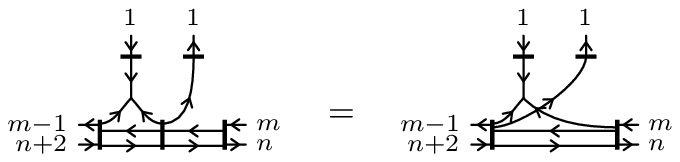}
\end{align}

\subsection*{weight $(m,n)\to (m,n)$}
The twisted diagram is
\begin{align}
\label{eq:spider_Ub1U3_twist}
&\eqfig{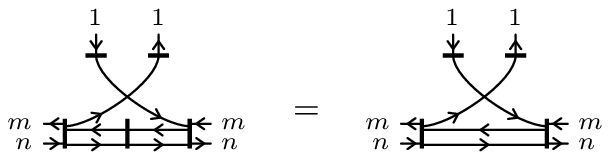}\\
\label{eq:spider_Ub3U1_twist}
&\eqfig{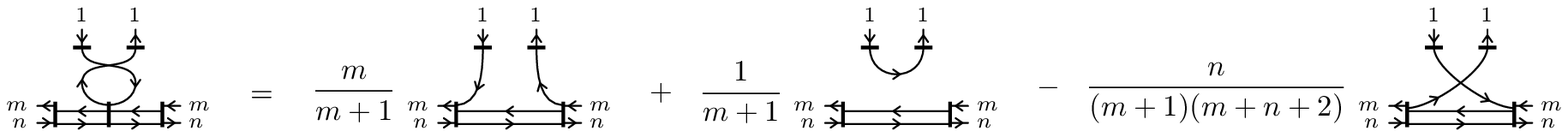}\\
\label{eq:spider_Ub2U2_twist}
&\eqfig{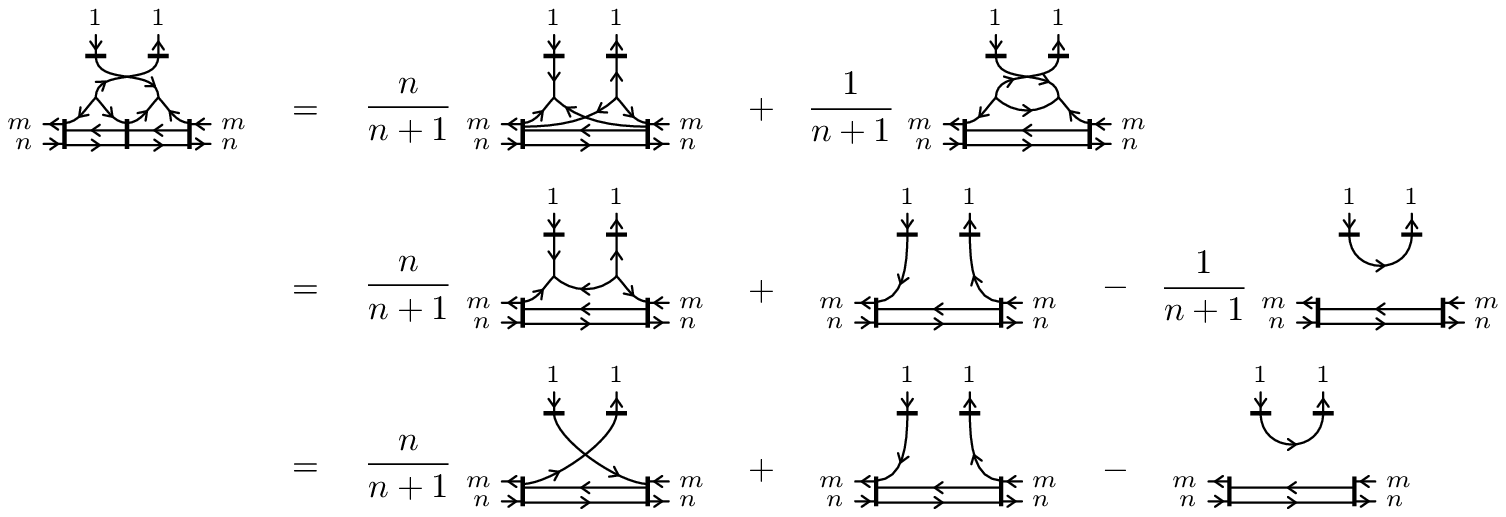}
\end{align}
The untwisted diagram is
\begin{align}
&\label{eq:spider_U1Ub3}
\eqfig{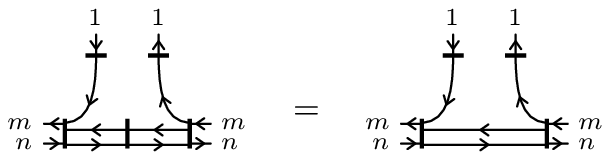}\\
&\label{eq:spider_U3Ub1}
\eqfig{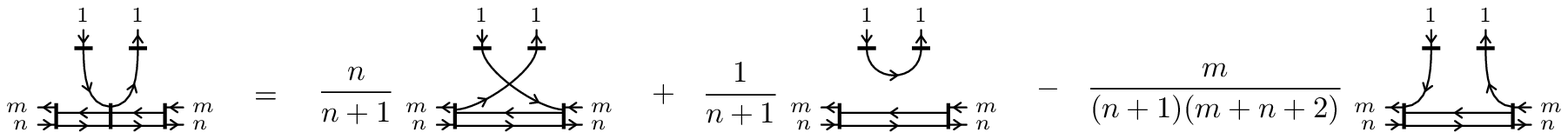}\\
&\label{eq:spider_U2Ub2}
\eqfig{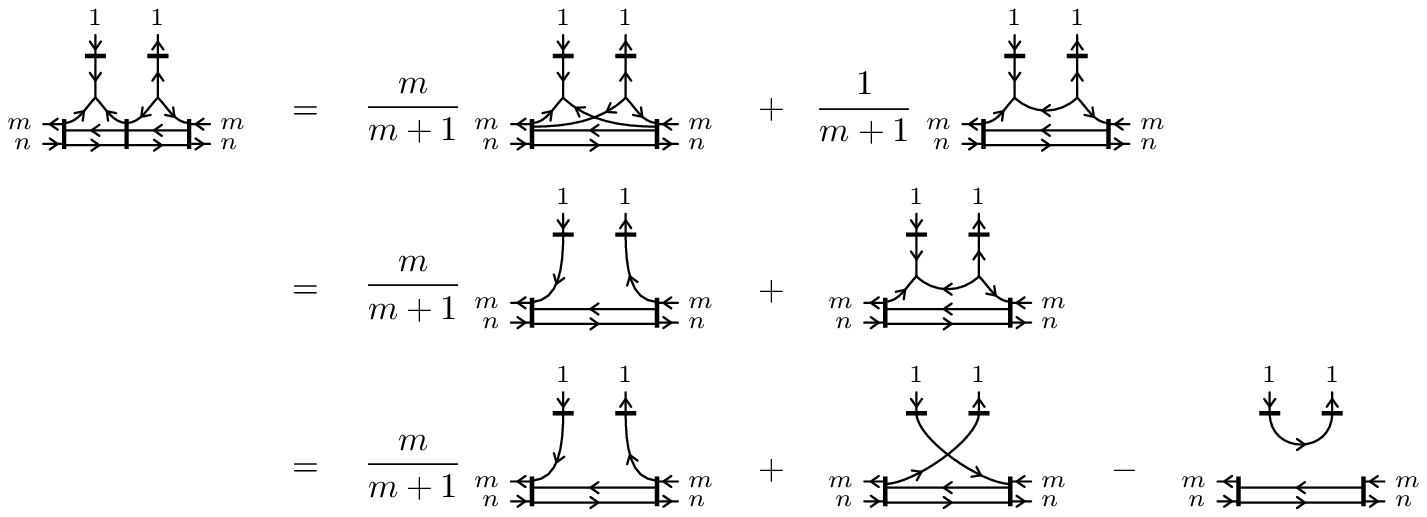}
\end{align}

\subsection*{weight $(m,n)\to (m-2,n+1)$}
The twisted diagram is
\begin{align}
\label{eq:spider_Ub3U2_twist}
&\eqfig{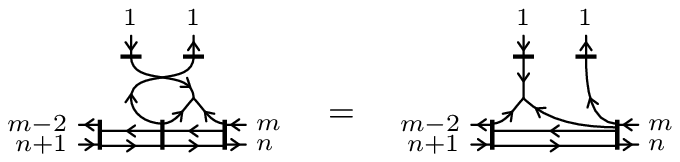}
\end{align}
The untwisted diagram is
\begin{align}
&\label{eq:spider_U2Ub3}
\eqfig{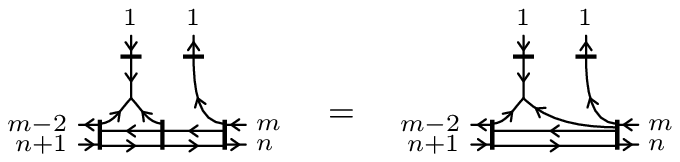}
\end{align}

\subsection*{weight $(m,n)\to (m+1,n-2)$}
The twisted diagram is
\begin{align}
\label{eq:spider_Ub2U3_twist}
&\eqfig{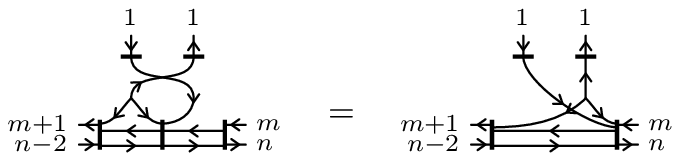}
\end{align}
The untwisted diagram is
\begin{align}
&\label{eq:spider_U3Ub2}
\eqfig{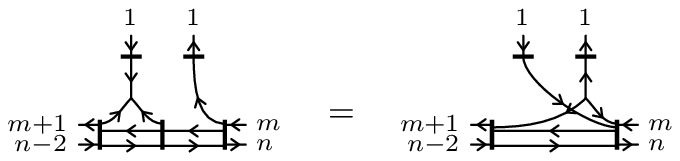}
\end{align}

\subsection*{weight $(m,n)\to (m-1,n-1)$}
The twisted diagram is
\begin{align}
\label{eq:spider_Ub3U3_twist}
&\eqfig{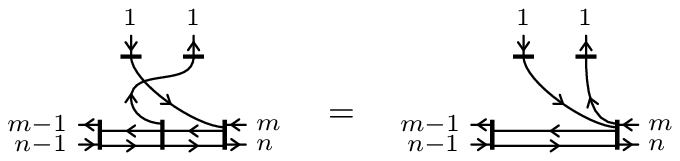}
\end{align}
The untwisted diagram is
\begin{align}
&\label{eq:spider_U3Ub3}
\eqfig{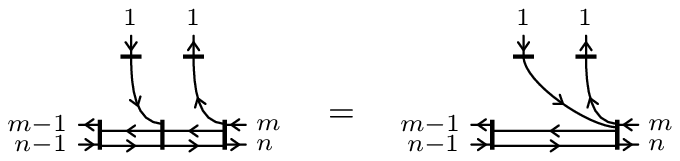}
\end{align}

\end{appendix}


\begin{thebibliography}{99}

\bibitem{Archer:1992kc} 
  F.~J.~Archer,
  \emph{``The U-q-sl(3) 6j symbols and state sum model,''}
  Phys.\ Lett.\ B {\bf 295}, 199 (1992).

\bibitem{1997q.alg....12003K} G.~Kuperberg, \emph{``Spiders for rank 2 Lie algebras,''} Comm. Math. Phys. {\bf 180}, 109--151, (1996).  



\bibitem{Moore:1988qv} 
  G.~W.~Moore and N.~Seiberg,
  \emph{``Classical and Quantum Conformal Field Theory,''}
  Commun.\ Math.\ Phys.\  {\bf 123}, 177 (1989).
\bibitem{AlvarezGaume:1989aq} 
  L.~Alvarez-Gaume, C.~Gomez and G.~Sierra,
  \emph{``Duality and Quantum Groups,''}
  Nucl.\ Phys.\ B {\bf 330}, 347 (1990).
\bibitem{Gerasimov:1990fi} 
  A.~Gerasimov, A.~Morozov, M.~Olshanetsky, A.~Marshakov and S.~L.~Shatashvili,
  \emph{``Wess-Zumino-Witten model as a theory of free fields,''}
  Int.\ J.\ Mod.\ Phys.\ A {\bf 5}, 2495 (1990).
  

\bibitem{Smirnov:2009fs} 
  A.~Smirnov,
  \emph{``Notes on Chern-Simons Theory in the Temporal Gauge,''}
  arXiv:0910.5011 [hep-th].
\bibitem{Mironov:2011ym} 
  A.~Mironov, A.~Morozov and A.~Morozov,
  \emph{``Character expansion for HOMFLY polynomials. I. Integrability and difference equations,''}
  arXiv:1112.5754 [hep-th].
\bibitem{Mironov:2011aa} 
  A.~Mironov, A.~Morozov and A.~.Morozov,
  \emph{``Character expansion for HOMFLY polynomials. II. Fundamental representation. Up to five strands in braid,''}
  JHEP {\bf 1203}, 034 (2012)
  arXiv:1112.2654 [math.QA].
\bibitem{Itoyama:2012qt} 
  H.~Itoyama, A.~Mironov, A.~Morozov and A.~.Morozov,
  \emph{``Character expansion for HOMFLY polynomials. III. All 3-Strand braids in the first symmetric representation,''}
  Int.\ J.\ Mod.\ Phys.\ A {\bf 27}, 1250099 (2012)
  arXiv:1204.4785 [hep-th].
\bibitem{Itoyama:2012re} 
  H.~Itoyama, A.~Mironov, A.~Morozov and A.~.Morozov,
  \emph{``Eigenvalue hypothesis for Racah matrices and HOMFLY polynomials for 3-strand knots in any symmetric and antisymmetric representations,''}
  Int.\ J.\ Mod.\ Phys.\ A {\bf 28}, 1340009 (2013)
  arXiv:1209.6304 [math-ph].
\bibitem{Anokhina:2013wka} 
  A.~Anokhina, A.~Mironov, A.~Morozov and A.~.Morozov,
  \emph{``Colored HOMFLY polynomials as multiple sums over paths or standard Young tableaux,''}
  Adv.\ High Energy Phys.\  {\bf 2013}, 931830 (2013)
  arXiv:1304.1486 [hep-th].
\bibitem{Nawata:2013ppa} 
  S.~Nawata, P.~Ramadevi and Zodinmawia,
  \emph{``Multiplicity-free quantum $6j$-symbols for $U_q(\mathfrak{sl}_N)$,''}
  Lett.\ Math.\ Phys.\  {\bf 103}, 1389 (2013)
  arXiv:1302.5143 [hep-th].


\bibitem{KiriResh1989}
A.N.Kirillov, N.Yu.Reshetikhin, \emph{``Representations of the algebra Uq(sl(2)), $q$-orthogonal polynomials and invariants of Links,''} In: Kohno, T. (ed.) New Developments in the Theory of Knots. World Scientific, Singapore (1989)
\bibitem{AlvarezGaume:1988vr} 
  L.~Alvarez-Gaume, C.~Gomez and G.~Sierra,
  \emph{``Quantum Group Interpretation of Some Conformal Field Theories,''}
  Phys.\ Lett.\ B {\bf 220}, 142 (1989).
\bibitem{Itoyama:1989mw} 
  H.~Itoyama and A.~Sevrin,
  \emph{``Braiding Matrices Of Conformal Blocks And Coset Models,''}
  Int.\ J.\ Mod.\ Phys.\ A {\bf 5}, 211 (1990).
\bibitem{Balog:1990dn} 
  J.~Balog, L.~Dabrowski and L.~Feher,
  \emph{``Classical R Matrix and Exchange Algebra in {WZNW} and Toda Theories,''}
  Phys.\ Lett.\ B {\bf 244}, 227 (1990).
\bibitem{Itoh:1990bi} 
  T.~Itoh and Y.~Yamada,
  \emph{``Explicit exchange relations in SL(2) Wess-Zumino-Novikov-Witten model,''}
  Int.\ J.\ Mod.\ Phys.\ A {\bf 6}, 3283 (1991).
\bibitem{Itoh:1992sq} 
  T.~Itoh,
  \emph{``Exchange relations in Wess-Zumino-Novikov-Witten model and quantum groups,''}
  TMUP-HEL-9202.
\bibitem{Smithies1995} 
	Ciudy Rae Smithies,
	\emph{``The $q$-deformed algebras $su(n)_{q}$ and their applications,''} thesis, University of Canterbury. Physics, (1995).


\bibitem{Wakimoto:1986gf} 
  M.~Wakimoto,
  \emph{``Fock representations of the affine lie algebra A1(1),''}
  Commun.\ Math.\ Phys.\  {\bf 104}, 605 (1986).
\bibitem{KT1998}
	G. Kuroki and T. Takebe,
	\emph{``Bosonization and Integral Representation of Solutions of the Knizhnik-Zamolodchikov-Bernard Equations''}, Comm. Math. Phys. {\bf 204}, 587, (1999)
	arXiv:math/9809157v2.


\bibitem{fuchs:book}
J\"urgen Fuchs,~
\emph{``Affine Lie Algebras and Quantum Groups,''} Cambridge University Press, (1995)
\bibitem{yamadabook2006} 
	Y. Yamada,
	\emph{``Introduction to Conformal Field Theories,''} Baifukan (2006), (Japanese).
\bibitem{AoKi:1994}
	K. Aomoto, M. Kita,
	\emph{``Theory of Hypergeometric Functions,''} Springer, (2011).
\end{thebibliography}
\end{document}